\documentclass{aa} 
\usepackage{graphicx, subfigure}
\usepackage[graphicx]{realboxes}
\usepackage{appendix}
\usepackage{xcolor, soul}
\sethlcolor{yellow}
\usepackage{txfonts}
\usepackage{eqnarray}
\usepackage{adjustbox}
\usepackage{enumitem}
\usepackage{natbib}
\usepackage{amsmath} 
\usepackage{hyperref} 

\definecolor{myblue}{RGB}{0, 160, 240} 
\definecolor{mygreen}{RGB}{0, 180, 0}

\usepackage{hyperref}
\hypersetup{colorlinks = true, citecolor = {blue}, linkcolor={blue}, urlcolor={blue}}

\begin{document} 

\title{The Milky Way Cepheid Leavitt law  based on Gaia DR2 parallaxes of companion stars and host open cluster populations}

\author{Louise Breuval \inst{1}, 
Pierre Kervella \inst{1}, 
Richard I. Anderson \inst{2}, 
Adam G. Riess \inst{3, 4}, 
Fr\'ed\'eric Arenou \inst{5}, 
Boris Trahin \inst{1}, \\
Antoine M\'erand \inst{2}, 
Alexandre Gallenne \inst{6, 7, 8, 9},  
Wolfgang Gieren \inst{7}, 
Jesper Storm \inst{10}, 
Giuseppe Bono \inst{11, 12}, \\
Grzegorz Pietrzyński \inst{7, 8}, 
Nicolas Nardetto \inst{6}, 
Behnam Javanmardi \inst{1},
Vincent Hocd\'e \inst{6}  }

\institute{LESIA, Observatoire de Paris, Universit\'e PSL, CNRS, Sorbonne Universit\'e, Univ. Paris Diderot, Sorbonne Paris Cit\'e, 5 place Jules Janssen, 92195 Meudon, France. \email{louise.breuval@obspm.fr}
        \and European Southern Observatory, Karl-Schwarzschild-Str. 2, 85748 Garching, Germany
        \and Space Telescope Science Institute, 3700 San Martin Drive, Baltimore, MD 21218, USA
        \and Department of Physics and Astronomy, Johns Hopkins University, Baltimore, MD 21218, USA
        \and GEPI, Observatoire de Paris, Universit\'e PSL, CNRS, 5 Place Jules Janssen, 92190 Meudon, France
        \and Universit\'e C\^ote d'Azur, Observatoire de la C\^ote d'Azur, CNRS, Laboratoire Lagrange, France
        \and Universidad de Concepci\'on, Departamento de Astronom\'ia, Casilla 160-C, Concepci\'on, Chile
        \and Nicolaus Copernicus Astronomical Centre, Polish Academy of Sciences, Bartycka 18, 00-716 Warszawa, Poland
        \and Unidad Mixta Internacional Franco-Chilena de Astronom\'ia (CNRS UMI 3386), Departamento de Astronom\'ia, Universidad de Chile, Camino El Observatorio 1515, Las Condes, Santiago, Chile
        \and Leibniz-Institut f{\"u}r Astrophysik Potsdam (AIP), An der Sternwarte 16, 14482 Potsdam, Germany
        \and    Department of Physics, Universit\`a di Roma Tor Vergata, via della Ricerca Scientifica 1, I-00133 Roma, Italy
        \and INAF-Osservatorio Astronomico di Roma, via Frascati 33, I-00040 Monte Porzio Catone, Italy
        }

\date{Received 11 June 2020 / Accepted 9 September 2020 }

\abstract
        {}
        {Classical Cepheids provide the foundation for the empirical extragalactic distance ladder. Milky Way Cepheids are the only stars in this class accessible to trigonometric parallax measurements. However, the parallaxes of Cepheids from the second Gaia data release (GDR2) are affected by systematics because of the absence of chromaticity correction, and occasionally by saturation.         }
        {As a proxy for the parallaxes of 36 Galactic Cepheids, we adopt either the GDR2 parallaxes of their spatially resolved companions or the GDR2 parallax of their host open cluster. This novel approach allows us to bypass the systematics on the GDR2 Cepheids parallaxes that is induced by saturation and variability. We adopt a GDR2 parallax zero-point (ZP) of -0.046 mas with an uncertainty of 0.015 mas that covers most of the recent estimates. }
        {We present new Galactic calibrations of the Leavitt law in the $V$, $J$, $H$, $K_S$, and Wesenheit $W_H$ bands. We compare our results with previous calibrations based on non-Gaia measurements and compute a revised value for the Hubble constant anchored to Milky Way Cepheids.     }
        {From an initial Hubble constant of $76.18 \pm 2.37$ km\,s$^{-1}$\,Mpc$^{-1}$ based on parallax measurements without Gaia, we derive a revised value by adopting companion and average cluster parallaxes in place of direct Cepheid parallaxes, and we find $H_0 = 72.8 \pm 1.9$ (statistical + systematics) $\pm$ 1.9 (ZP) km\,s$^{-1}$\,Mpc$^{-1}$ when all Cepheids are considered and $H_0 = 73.0 \pm 1.9$ (statistical + systematics) $\pm$ 1.9 (ZP) km\,s$^{-1}$\,Mpc$^{-1}$ for fundamental mode pulsators only. }


\titlerunning{The Milky Way Cepheid Leavitt law based on Gaia DR2 parallaxes}
\authorrunning{Breuval et al.}
\maketitle

\section{Introduction}\label{sec:intro}

Classical Cepheids (CCs) have a historical  major importance among variable stars because of the simple correlation between the pulsation period and  intrinsic luminosity, also called the Leavitt law or the period--luminosity (PL) relation \citep{1908AnHar..60...87L, 1912HarCi.173....1L}. However, after more than a century of active research, the absolute calibration of the Leavitt law is still unsatisfactory because of the lack of precise and direct distance measurements for a sizeable sample of these stars. A careful calibration of this relation and especially of its zero-point is fundamental as it is used to establish extragalactic distances and to derive the expansion rate of the Universe,  the Hubble constant $H_0$. The determination of $H_0$ from the Cosmic Microwave Background (CMB) based on the standard $\Lambda$ Cold Dark Matter ($\Lambda$CDM) model \citep{2018arXiv180706209P} is currently found to be in $\sim 5\sigma$ tension with the empirical or direct distance ladder measurements \citep{2019NatRP...2...10R}. This tension may have important implications in cosmology, and may even point toward new physics beyond $\Lambda$CDM \citep{2019NatAs...3..891V}.

Calibrating the Leavitt law  requires  independent and accurate distance measurement for a sample of CCs. Unfortunately, Gaia's second data release (hereafter GDR2) contains a number of systematic effects that may reduce the precision of the parallaxes of CCs \citep{2018AA...616A...1G}. First, CCs are bright stars, so a small number with $G<6$ mag are affected by saturation, making their parallaxes unreliable. In addition, CC colors cycle through many variations during the parallax cycle; the effective temperature of a Cepheid changes on average by 1000 K over a full pulsation cycle \citep{2018AA...616A..82P}, which means $\sim$0.5 mag in optical bands, so this may add additional noise to their astrometry due to the chromaticity of the PSF. Future Gaia data releases are expected to include chromaticity corrections for variable stars and incorporate a better model of the PSF to deal with saturation. While recent analyses of Gaia DR2 parallaxes for CCs with G>6 mag do not appear to be affected by excess noise (an indicator of poor quality for GDR2 astrometric data) \citep{2018AA...619A...8G, 2018ApJ...861..126R, 2017A&A...605A..79G, Clementini2019}, it is important to pursue alternative approaches to extract parallaxes from Gaia DR2 for CCs that are insensitive to these systematics.

Even in the absence of systematic errors, the use of open cluster parallaxes for the CCs they host can provide enhanced precision over the use of a single CC parallax. Because open cluster parallaxes are based on many stars, the increased precision from averaging and the ability to reject outliers for stars in astrometric binaries is extremely valuable.

In the present paper our aim is to calibrate the Milky Way (MW) Cepheid Leavitt law using stars that are not affected by these issues and to benefit from the gain in precision afforded by cluster average parallaxes. In Sect.~\ref{sec:sample} we introduce our sample of stars and their associated parallaxes and photometry. In Sect.~\ref{subsec:Leavitt_law} we derive calibrations of the Leavitt law in various bands. Then in Sect.~\ref{subsec:comparisonHST} we compare our GDR2 parallaxes with the corresponding expected parallaxes from Hubble Space Telescope (HST) measurements, and in Sect. \ref{subsec:H0} we derive a value for the Hubble constant anchored to Milky Way Cepheids.

\section{Sample} \label{sec:sample}

We consider two sets of parallaxes: one based on Cepheid companions and one based on average cluster parallaxes. The benefits of these samples are flux and color constancy (companions and clusters) and averaging over a large sample (clusters).

        \subsection{Parallaxes of Cepheid resolved companions}\label{subsec:companions}
        
\begin{figure}[]
\centering
\includegraphics[height=8cm]{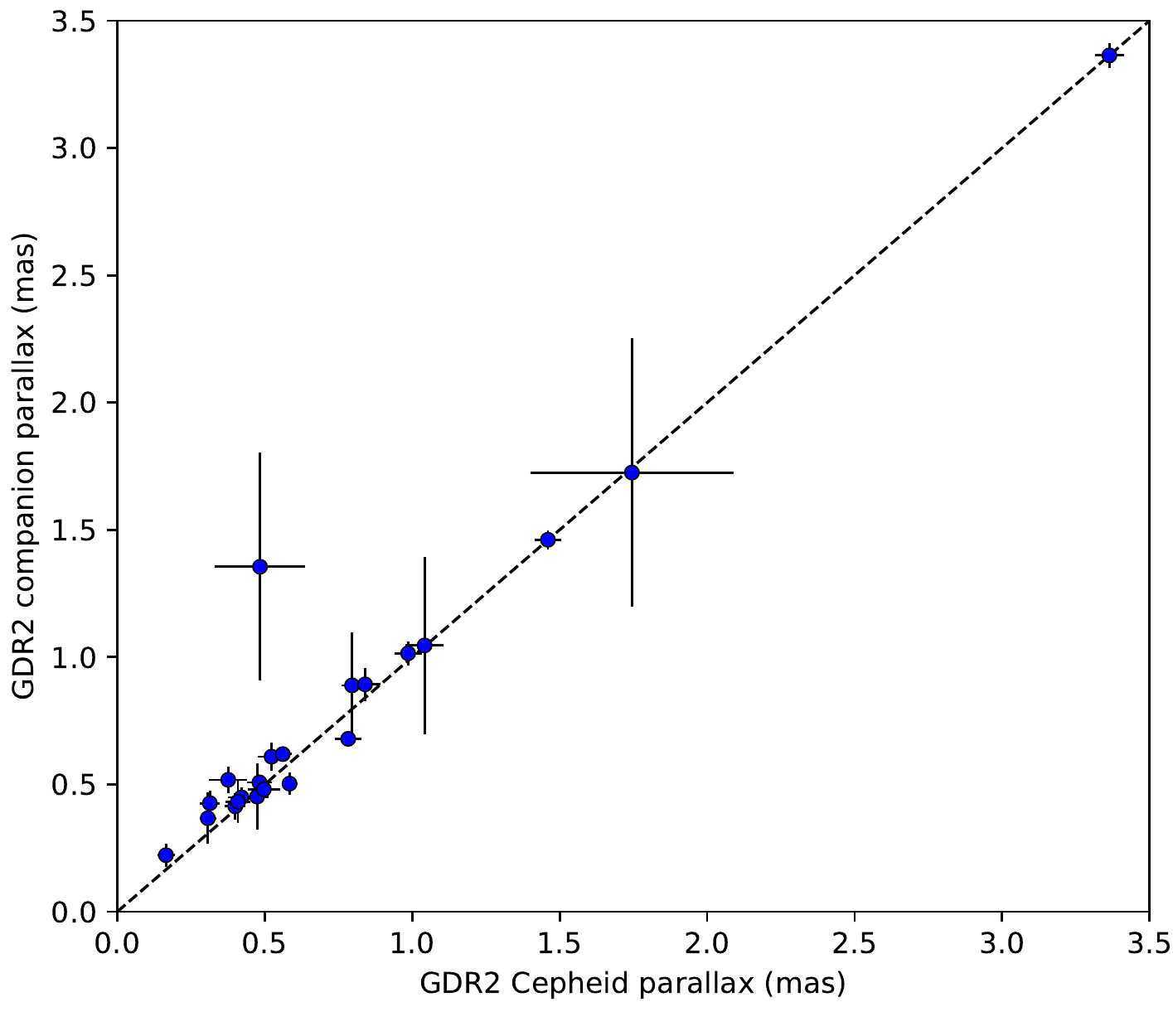}
\caption{GDR2 parallaxes of our sample of companions as a function of the corresponding GDR2 Cepheid parallax. The dashed line corresponds to the identity line.}
\label{fig:cep_comp}
\end{figure}

Recently, \citet{2019A&A...623A.117K} presented a sample of 28 Galactic Cepheids that are members of gravitationally bound and spatially resolved stellar systems. In these systems Cepheid companions are photometrically stable stars and their GDR2 parallaxes are therefore not affected by such a strong chromatic effect as Cepheids. As the CCs and their companions share the same parallax (their relative distance is negligible compared to the distance to Gaia), the GDR2 parallaxes of the companions provide a natural proxy for those of the CCs. The companions' parallaxes are precise within 15\%, on average. A comparison between direct GDR2 Cepheid parallaxes and the corresponding GDR2 companion parallaxes is displayed in Fig. \ref{fig:cep_comp}.

The angular separation between the CCs and their companions is in most cases larger than 10 arcsec, which is large enough to prevent flux contamination, given the brightness of the CCs. At 10" separation for stars hundreds to thousands of parsec distant there is no expected effect of orbital motion on parallax or proper motion measurements: the parallaxes of the CCs and companions are not sensitive to the binarity of these wide systems.

The GDR2 astrometry is generally of poor quality for very bright stars ($G <$ 6 mag), due to calibration issues and saturation \citep{2018ApJ...861..126R, 2019RNAAS...3f..79D, 2019arXiv190609827L}. This occurs independently of the chromaticity issue raised previously, whether the star is variable or not. While several Cepheids  in our sample are close to this limit, with an average $G$ magnitude of 8 mag, their companions are on average 7 mag fainter than their parent Cepheids. The companions are therefore not as affected as CCs by the saturation issue and they are far away from the sensitivity limit. They consequently belong to the best dynamical range for \textit{Gaia}. 

For a given Cepheid, when more than one companion was found by \citet{2019A&A...623A.117K}, we selected the companion with the smallest uncertainty on its parallax. This selection was performed for CV Mon, SY Nor, U Sgr, and V350 Sgr.

Various quality indicators are introduced in the second release of Gaia data, such as the re-normalized unit weight error (RUWE, noted $\varrho$ in the following). It is particularly pertinent because it evaluates the quality of the parallax of a star compared to other stars of the same type. This parameter is defined by \citet{Lindegren2018} as  
\begin{equation}
 \varrho = \frac{\mathrm{UWE}}{u_0 (G, C)} 
 ,\end{equation}
where $\mathrm{UWE} = \sqrt{\chi^2 / (N-5)}$  is the unit weight error and $u_0$ is an empirical normalization  factor that is not directly available in the Gaia release, but which can be computed from the lookup table on the ESA DR2 \textit{Known issues} web page\footnote{\url{https://www.cosmos.esa.int/web/gaia/dr2-known-issues}}. Following \citet{Lindegren2018}, we estimate that a parallax is reliable if $\varrho<1.4$. The Table \ref{table:plx_companions} gives the RUWE for the Cepheids and the companions in our sample.

We note that some CCs from the \citet{2019A&A...623A.117K} sample have no valid GDR2 parallax ($\delta$ Cep, R Cru, $\alpha$ UMi), while all companions have a valid parallax. In the initial \citet{2019A&A...623A.117K} sample of 28 Cepheids, five of them have $\varrho>1.4$, while only two companions are in this case, R Cru and V1046 Cyg, with $\varrho = 2.80$ and $1.51,$ respectively. We exclude these two stars from the sample of companions in order to keep accurate parallaxes only. The star CE Cas B is a particular case because its companion CE Cas A is also a Cepheid. We exclude both stars from our sample as a precaution.

\begin{table*}[]
\small
\caption{Sample of Cepheids in resolved binary systems taken from \citet{2019A&A...623A.117K}: parameters of the Cepheids and of their stable companions. The symbol $\varrho$ is the RUWE quality indicator from GDR2 and ($\ast$) indicates that $\varrho > 1.4$.}
\centering
\begin{tabular}{l c l c c | l c l c c}
\hline
\hline
Cepheid & $\varpi_{\rm GDR2}$   & $\varrho$     & $G$ & $BP$-$RP$       & Companion (GDR2)        & $\varpi_{\rm GDR2}$   & $\varrho$     & $G$ & $BP$-$RP$ \\
                & \tiny (mas)                           &                       & \tiny (mag)     & \tiny (mag) &                                         & \tiny (mas)                             &                       & \tiny (mag) & \tiny (mag) \\
\hline
DF Cas          & $0.307_{\pm 0.028}$& 0.98             & 10.43 & 1.50  & 465719182408531072   & $0.367_{\pm 0.101}$  & 1.12 & 17.26       & 2.02    \\
CM Sct          & $0.376_{\pm 0.065}$& 1.02             & 10.51 & 1.81  & 4253603428053877504 & $0.518_{\pm 0.051}$  & 0.95 & 14.73 & 1.49  \\
EV Sct          & $0.497_{\pm 0.054}$& 1.05             & 9.62  & 1.63  & 4156513016572003840 & $0.481_{\pm 0.034}$  & 1.03 & 13.62 & 0.88    \\
TV CMa          & $0.314_{\pm 0.034}$& 0.94             & 10.08 & 1.69  & 3044483895574944512 & $0.426_{\pm 0.049}$  & 1.12 & 15.77 & 1.18   \\
V532 Cyg                & $0.561_{\pm 0.032}$& 0.86             & 8.67  & 1.44  & 1971721839529622272 & $ 0.619_{\pm 0.027}$ & 0.93 & 14.67 & 1.04    \\
V950 Sco                & $0.840_{\pm 0.052}$& 1.09             & 7.05  & 1.05  & 5960623340819000192 & $0.893_{\pm 0.065}$  & 1.00 & 15.28 & 1.03   \\
V350 Sgr                & $0.986_{\pm 0.047}$& 0.92             & 7.25  & 1.26  & 4080121319521641344 & $1.015_{\pm 0.048}$  & 0.96 & 12.27 & 0.50    \\
VW Cru          & $0.783_{\pm 0.045}$& 0.98             & 9.01  & 1.85  & 6053622508133367680 & $0.679_{\pm 0.028}$  & 0.96 & 14.07 & 1.18    \\
AX Cir          & $1.745_{\pm 0.345}$& 10.3 $\ast$      & 5.63  & 1.15  & 5874031027625742848 & $1.725_{\pm 0.527}$  & 1.13 & 19.82 & 1.65   \\
$\delta$ Cep    &  -                        & 20.9 $\ast$       & -              & -         & 2200153214212849024 & $3.364_{\pm 0.049}$  & 0.85 & 6.28 & -0.02    \\
CV Mon          & $0.482_{\pm 0.041}$& 1.16             & 9.61  & 1.78  & 3127142327895572352 & $0.508_{\pm 0.025}$  & 1.03 & 13.49 & 1.03    \\
QZ Nor          & $0.474_{\pm 0.038}$& 1.01             & 8.58  & 1.18  & 5932565899990412672 & $0.452_{\pm 0.130}$  & 1.02 & 17.93 & 1.29    \\
V659 Cen                & $0.484_{\pm 0.154}$& 4.52 $\ast$      & 6.39  & 1.04  & 5868451109212716928 & $1.355_{\pm 0.448}$  & 1.00 & 19.69 & 2.50     \\
CS Vel          & $0.165_{\pm 0.030}$& 1.01             & 11.10         & 1.83  & 5308893046071732096 &  $0.222_{\pm 0.045}$  & 0.95 & 16.20 & 1.08    \\
RS Nor          & $0.421_{\pm 0.046}$& 0.99             & 9.49          & 1.73  & 5932812740361508736 & $0.449_{\pm 0.038}$  & 1.06 & 14.55  & 0.97  \\
X Cru           & $0.523_{\pm 0.046}$& 0.97             & 8.07  & 1.28  & 6059762524642419968 & $0.609_{\pm 0.056}$  & 1.16 & 16.04  & 1.11   \\
AW Per          & $1.042_{\pm 0.064}$& 1.06             & 7.05  & 1.97  & 174489098011144960   & $1.046_{\pm 0.348}$  & 1.07 & 17.42  & 1.54    \\
U Sgr           & $1.460_{\pm 0.045}$& 1.06             & 6.35  & 1.54  & 4092905203841177856 & $1.461_{\pm 0.038}$  & 0.90 & 11.14  & 0.67   \\
ER Car          & $0.796_{\pm 0.035}$& 0.95             & 6.61  & 1.08  & 5339394048386734336 & $0.889_{\pm 0.208}$  & 1.09 & 18.44 & 1.37   \\
SX Vel          & $0.409_{\pm 0.041}$& 1.00             & 7.97  & 1.24  & 5329838158460399488 & $0.432_{\pm 0.083}$  & 0.95 & 17.02 & 1.13    \\
SY Nor          & $0.400_{\pm 0.035}$& 1.10             & 8.97  & 1.83  & 5884729035245399424 & $0.414_{\pm 0.053}$  & 1.28 & 12.10  & 0.88    \\
RS Pup          & $0.584_{\pm 0.026}$& 0.97             & 6.46  & 1.88  & 5546476755539995008 & $0.503_{\pm 0.045}$  & 1.00 & 16.25  & 1.28     \\
\hline
\end{tabular}
\label{table:plx_companions}
\end{table*}

The star $\alpha$ UMi is extremely bright, with $K \approx 0.5$ mag. Therefore, measuring accurate photometry for this star is particularly challenging. It has no valid parallax in GDR2 and appears saturated in most catalogs \citep{2006AJ....131.1163S}. The only accurate average magnitudes based on several pulsation cycles were found in the AAVSO database that provides $J=0.93 \pm 0.01$ mag and $H=0.67 \pm 0.01$ mag in the UKIRT system. Additionally, the uncertain pulsation mode and the age difference between the Cepheid and its companion raise questions concerning the properties of $\alpha$ UMi and whether it should be included in PL relation fits \citep{2018AA...611L...7A, 2018ApJ...853...55B, 2018AA...619A...8G}. We decided to exclude this star from our sample. 

Finally, this selection results in a sample of 22 GDR2 parallaxes of Cepheids resolved companions, listed in Table \ref{table:plx_companions}.

        \subsection{Parallaxes of Cepheids in open clusters}
        \label{subsec:clusters}

Open clusters (OCs) contain a significant number of stars located at the same distance and are numerous in the Milky Way. Therefore, identifying Cepheids in OCs allows us to estimate their distances, with an important gain in precision by taking the average over a population compared to individual parallax measurements. 

We performed a cross-match between the \citet{2019AA...625A..14R} reclassification of GDR2 Cepheids and the \citet{2018AA...618A..93C} catalog of Milky Way Open Clusters. This catalog provides parallaxes for 1229 OCs, computed as the median GDR2 parallax of their member stars. Our comparison is based on five membership constraints:   separation $\theta$, parallax $\varpi$, proper motion $\mu_{\alpha}^{\ast}$ and $\mu_{\delta}$, and age. 

\begin{figure}[]
\centering
\includegraphics[height=8cm]{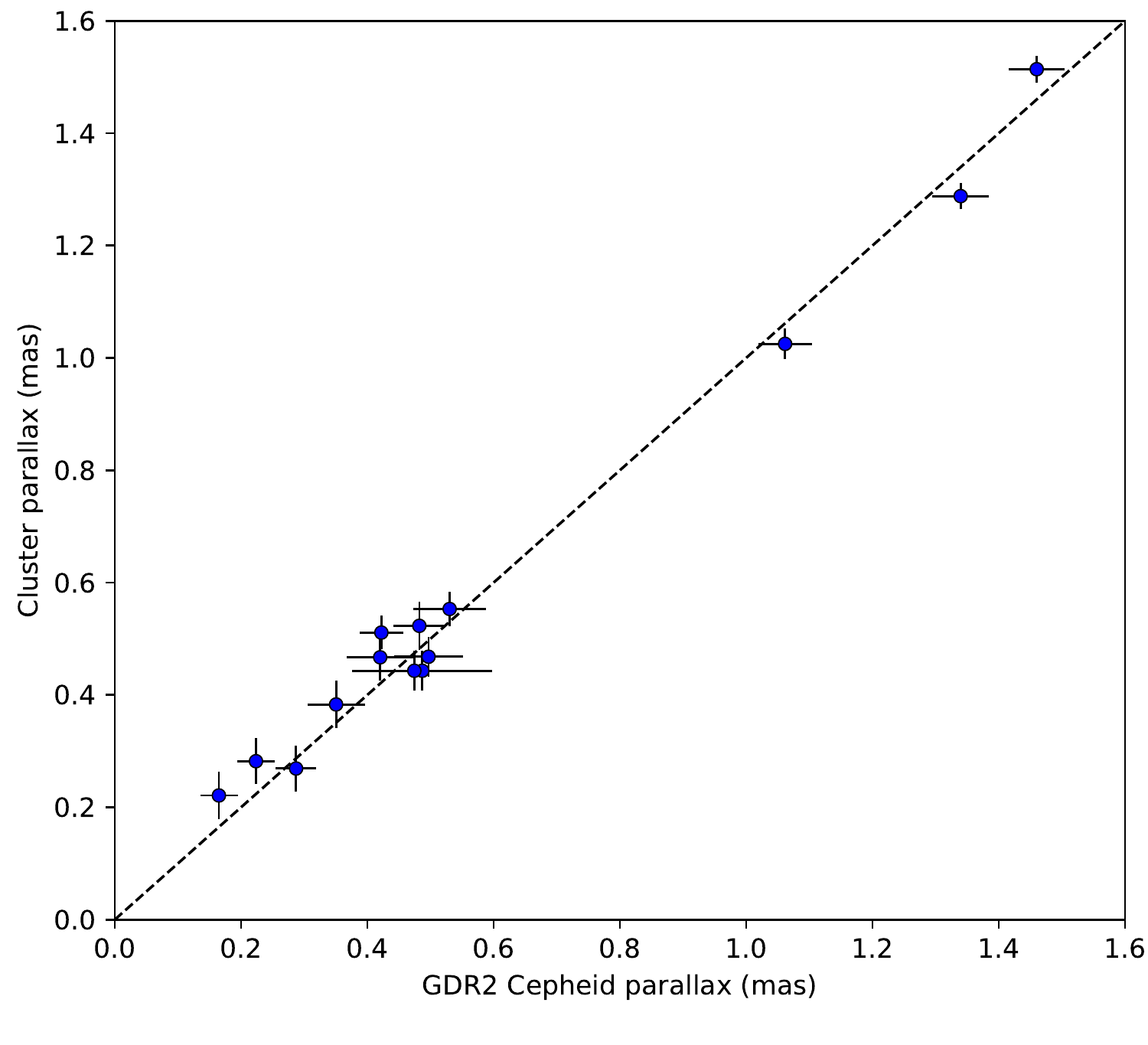}
\caption{Parallaxes of our sample of cluster Cepheids from \citet{2018AA...618A..93C} with revised uncertainties, as a function of the corresponding GDR2 Cepheid parallax. The dashed line corresponds to the identity line.}
\label{fig:cep_clust}
\end{figure}

\begin{table*}[]
\small
\caption{Sample of cluster Cepheids found by our cross-match selection and in the literature. Full circles stand for an agreement smaller than 1$\sigma$ between the Cepheid and the cluster parameters.}
\centering
\begin{tabular}{l l | c c c c c c | c c c c c c}
\hline
\hline
                &                       & \multicolumn{6}{c|}{Cross-match} & \multicolumn{5}{c}{Cluster parameters} \\
\hline
Cepheid         & Cluster               & $\varpi$      & $\mu_{\alpha}^*$ & $\mu_{\delta}$ & Age  & Sep  & Ref       & $\varpi_{\rm CG18}$ & $N_{\rm memb}$ & $r_{50}$ & $<V_{\varpi}>$ & $\varpi_{\rm adopted}$  \\
        &               &       & &  &          & \tiny (arcmin) &  & \tiny (mas) & & \tiny (deg) & \tiny ($\rm \mu as^2$) & \tiny (mas) \\  
\hline 
CV Mon  & vdBergh 1     & $\bullet$     & $\bullet$     & $\bullet$     & $\bullet$       & 0.9  & a, b, c, d             & $0.523_{\pm 0.010}$ &  73  &  0.03 & 1741 & $0.523_{\pm 0.043}$ \\
S Nor   & NGC 6087      & $\bullet$     & $\bullet$     & $\bullet$     & $\bullet$       & 1.0  & a, b, c, e             & $1.025_{\pm 0.004}$ &  251  &  0.25 & 708 & $1.025_{\pm 0.027}$  \\
U Sgr   & IC 4725       & $\bullet$     & $\bullet$     & $\bullet$     & $\bullet$       & 2.1  & a, b, c                & $1.514_{\pm 0.003}$ &  516  &  0.26 & 563 & $1.514_{\pm 0.024}$   \\
V367 Sct        & NGC 6649      & $\bullet$     & $\bullet$     & $\bullet$         & $\bullet$     & 2.8     & a   & $0.467_{\pm 0.004}$ &  560  &  0.06 & 1689 & $0.467_{\pm 0.041}$    \\
V Cen   & NGC 5662      & $\bullet$     & $\bullet$     & $\bullet$     & $\bullet$       & 25   & a, b, c, f             & $1.288_{\pm 0.003}$ &  255  &  0.33 & 533 & $1.288_{\pm 0.023}$   \\
RS Ori  & FSR 0951      & $\bullet$     & $\bullet$     & $\bullet$     & 2.4 $\sigma$& 2.0         & PW          & $0.553_{\pm 0.004}$ &  195  &  0.16 & 697 & $0.553_{\pm 0.027}$   \\
CS Vel  & Ruprecht 79   & $\bullet$     & $\bullet$     & $\bullet$     & 2.5 $\sigma$    & 2.2  & g      & $0.221_{\pm 0.004}$ &  178  &  0.05 & 1720 & $0.221_{\pm 0.042}$   \\
DL Cas  & NGC 129       & $\bullet$     & $\bullet$     & 3.4 $\sigma$  & $\bullet$       & 3.4  & a, b, c                & $0.511_{\pm 0.002}$ &  392  &  0.17  & 904 & $0.511_{\pm 0.031}$ \\
EV Sct  & NGC 6664      & $\bullet$     & 1.1 $\sigma$  & $\bullet$     & $\bullet$       & 2.4  & a              & $0.468_{\pm 0.004}$ &  237  &  0.10 & 1215 & $0.468_{\pm 0.035}$     \\
V340 Nor        & NGC 6067      & $\bullet$     & 1.3 $\sigma$  & $\bullet$         & $\bullet$     & 0.9  & a, c           &  $0.443_{\pm 0.002}$ &  995  &  0.11 & 1263 & $0.443_{\pm 0.036}$  \\
CF Cas  & NGC 7790      & $\bullet$     & 1.9 $\sigma$  & $\bullet$     & $\bullet$       & 1.3  & a, b, c        & $0.269_{\pm 0.004}$ &  200  &  0.06 & 1642 & $0.269_{\pm 0.041}$  \\
TW Nor  & Lyng\r{a} 6   & $\bullet$     & 2.0 $\sigma$  & $\bullet$     & 1.4 $\sigma$    & 0.6  & a, b, c, h             & $0.383_{\pm 0.006}$ &  79  &  0.06 & 1730 & $0.383_{\pm 0.042}$    \\
QZ Nor  & NGC 6067      & $\bullet$     & 1.2 $\sigma$  & 9 $\sigma$    & $\bullet$       & 18   & a              & $0.443_{\pm 0.002}$ &  955  &  0.11 & 1263 & $0.443_{\pm 0.036}$  \\
CG Cas  & Berkeley 58   &  $\bullet$    & 4.1 $\sigma$  & 2.0 $\sigma$  & 1.2 $\sigma$ & 5.5  & a, b              & $0.282_{\pm 0.004}$ &  142  &  0.06 & 1661 & $0.282_{\pm 0.041}$   \\
\hline
\end{tabular}
\label{table:plx_clusters}
\tablebib{(a): \citet{2013MNRAS.434.2238A}; (b): \citet{Chen2015}; (c): \citet{An2007}; (d): \citet{Turner1998}; (e): \citet{Turner1986}; (f): \citet{Turner1982}; (g): \citet{2010Ap&SS.326..219T}; (h): \citet{Majaess2011}; (PW): Present work. }
\end{table*}

Following \citet{2013MNRAS.434.2238A}, we start the search for potential cluster members by looking at the proximity in the sky: we selected all Cepheids located in a region of 10$r_{50}$ around each cluster (where $r_{50}$ is the radius containing half of the members) and we find a total of 2647 couples. For these couples we compared the parallaxes, the proper motions, and the ages of the two components. Since GDR2 parallaxes of Cepheids may be affected by systematics due to the absence of chromaticity correction, we account for this effect by including 20\% error in quadrature. The proper motions for Cepheids and open clusters are taken from \citet{2019AA...625A..14R} and \citet{2018AA...618A..93C}, respectively. The age of open clusters is provided by \citet{2013yCat..35580053K}, and the age for Cepheids is derived using period--age relations from \citet{2016AA...591A...8A}. 

We also searched in the literature for additional combinations and examined whether they satisfy our membership constraints. Some Cepheids are not present in the \citet{2019AA...625A..14R} reclassification, so they could not be found by means of our cross-match. \citet{2013MNRAS.434.2238A}   presented many of our couples and provided three additional combinations that verify our membership criteria: TW Nor, CV Mon, and V0367 Sct respectively in Lyng\r{a} 6, vdBergh 1, and NGC 6649. Other studies, such as \citet{An2007} and \citet{Chen2015}, also confirm most of our cluster memberships. Recently, \citet{Clark2015} and \citet{Lohr2018} identified new Cepheids as potential members of open clusters. However, no near-infrared (NIR) multi-epoch photometry is available for these Cepheids. Moreover, the \citet{Clark2015} starburst cluster VdBH 222 is not present in the \citet{2018AA...618A..93C} catalog. Therefore, we did not include them in our sample.

We find a total of 14 Cepheids that are candidate members of open clusters. They are listed in Table \ref{table:plx_clusters}, where filled circles stand for the agreement of a parameter at 1$\sigma$ or less. In this table is also provided the separation in arcmin between a Cepheid and the center of its host cluster. 

Due to the limited angular size of a cluster, parallaxes of the member stars of a same cluster are highly correlated. Uncertainties provided by \citet{2018AA...618A..93C} neglect this effect. Therefore, we revised the open cluster parallax uncertainties by including spatial correlations. We used the approach described in \citet{Lindegren2018b} and retrieved the spatial covariance $V_{\varpi}(\theta)$ of parallax errors on the ESA DR2 \textit{Known issues} web page\footnote{\url{https://www.cosmos.esa.int/web/gaia/dr2-known-issues}}. For each cluster, Table \ref{table:plx_clusters} provides the original \citet{2018AA...618A..93C} uncertainties, the number of member stars in each cluster, the cluster radius $r_{50}$, the averaged $V_{\varpi}(\theta)$, and  the adopted parallax $\varpi_{\rm adopted}$ with its revised uncertainty. After this correction the average precision of our cluster parallaxes is $\sim$ 8\%. 

The Cepheid QZ Nor is a particular case; located at 18 arcmin of NGC 6067, it is a peripherical member of this cluster. The 9$\sigma$ difference in $\mu_{\delta}$ could be explained by the fact that the Cepheid is leaving the cluster. This membership was identified by \citet{2013MNRAS.434.2238A} as bona fide. Moreover, QZ Nor is also present in the sample of companions found by \citet{2019A&A...623A.117K}: the stable star Gaia DR2 5932565899990412672 is located at 16'' (30 kau) from the Cepheid. Its GDR2 parallax of 0.452 $\pm$ 0.130 mas agrees particularly well with the 0.443 $\pm$ 0.036 mas parallax of NGC 6067 from \citet{2018AA...618A..93C}. Therefore, we decided to include this pair.

The cross-match also resulted in potential members that only have 2MASS single epoch photometry available. Since average magnitudes are preferred for the Leavitt law calibration, we discarded these pairs. In that case, we found V379 Cas, GU Nor, and XZ Car to be members of NGC 129, NGC 6067, and NGC 3496 respectively.

A comparison between direct GDR2 Cepheid parallaxes and the corresponding open cluster parallaxes from \citet{2018AA...618A..93C} is displayed in Fig. \ref{fig:cep_clust}. The field charts of each open cluster Cepheid are displayed in Figs. \ref{fields_1_of_3}, \ref{fields_2_of_3}, and \ref{fields_3_of_3} in the Appendix. 

        \subsection{Photometry}
        \label{subsec:photometry}
        
\begin{table*}[]
\caption{Final sample adopted, combining Cepheids with resolved companions and open cluster Cepheids. Parallaxes in the first part of the table are from GDR2 for the companions;  parallaxes in the second part are from \citet{2018AA...618A..93C} based on GDR2 with revised uncertainties. Reddenings $E(B-V)$ are taken from  the DDO database \citep{1995IBVS.4148....1F}, to which we applied a multiplicative factor of 0.94. Mean apparent magnitudes in $V$, $J$, $H$, $K_S$ bands are from the catalog compiled by \citet{2018AA...619A...8G}: $V$ band magnitudes are originally from \citet{2015AN....336...70M} and NIR magnitudes are converted in the 2MASS system with the original references provided in the last column. Apparent Wesenheit magnitudes on the WFC3 system ($m_H^W$) are also provided; their uncertainties include the photometric transformation errors.}
\begin{center}
\small
\begin{tabular}{l l c c c c c c c c}
\hline
\hline
Cepheid & $P$   & $\varpi^{\,(*)}$ & $E(B-V)$   & $m_V$         & $m_J$         & $m_H$   & $m_{K_S}$     &  $m_H^{W ~ \,(**)}$    & ref$_{\rm NIR}$ \\
                & \tiny (days)  & \tiny (mas)   &                       & \tiny (mag)     & \tiny (mag)   & \tiny (mag)   & \tiny (mag)   &  \tiny (mag) \\
\hline
 & \multicolumn{8}{c}{Sample of Cepheids with resolved companions}  \\
 \hline
\object{DF Cas} & 3.832  & 0.367$_{\pm 0.104}$ & 0.564$_{\pm 0.049}$ & 10.880$_{\pm 0.030}$ & 8.488$_{\pm 0.025}$ & 8.036$_{\pm 0.025}$ & 7.879$_{\pm 0.025}$ & 7.533$_{\pm 0.066}$ & G14 \\
\object{CM Sct} & 3.917  & 0.518$_{\pm 0.056}$ & 0.775$_{\pm 0.045}$ & 11.100$_{\pm 0.030}$ & 8.300$_{\pm 0.025}$ & 7.818$_{\pm 0.025}$ & 7.558$_{\pm 0.025}$ & 7.240$_{\pm 0.066}$  & G14    \\
\object{EV Sct} & 4.396 ${\star}$ & 0.481$_{\pm 0.040}$ & 0.623$_{\pm 0.015}$ & 10.130$_{\pm 0.030}$ & 7.608$_{\pm 0.008}$ & 7.184$_{\pm 0.008}$ & 7.018$_{\pm 0.008}$ & 6.658$_{\pm 0.061}$ & L92      \\
\object{TV CMa} & 4.670  & 0.426$_{\pm 0.054}$ & 0.574$_{\pm 0.029}$ & 10.590$_{\pm 0.030}$ & 8.022$_{\pm 0.008}$ & 7.582$_{\pm 0.008}$ & 7.364$_{\pm 0.008}$ & 7.048$_{\pm 0.061}$ & M11 \\
\object{V532 Cyg} & 4.675 ${\star}$ & 0.619$_{\pm 0.033}$ & 0.519$_{\pm 0.007}$ & 9.090$_{\pm 0.030}$ & 6.863$_{\pm 0.025}$ & 6.393$_{\pm 0.025}$ & 6.250$_{\pm 0.025}$ & 5.919$_{\pm 0.066}$ & 2MASS \\
\object{V950 Sco} & 4.814 ${\star}$ & 0.893$_{\pm 0.069}$ & 0.251$_{\pm 0.019}$ & 7.310$_{\pm 0.030}$ & 5.681$_{\pm 0.008}$ & 5.439$_{\pm 0.008}$ & 5.295$_{\pm 0.008}$ & 5.083$_{\pm 0.061}$ & G18 \\
\object{V350 Sgr} & 5.154  & 1.015$_{\pm 0.048}$ & 0.308$_{\pm 0.008}$ & 7.470$_{\pm 0.030}$ & 5.625$_{\pm 0.010}$ & 5.245$_{\pm 0.010}$ & 5.121$_{\pm 0.010}$ & 4.844$_{\pm 0.061}$ & W84 \\
\object{VW Cru} & 5.265  & 0.679$_{\pm 0.034}$ & 0.640$_{\pm 0.046}$ & 9.600$_{\pm 0.030}$ & 6.805$_{\pm 0.025}$ & 6.261$_{\pm 0.025}$ & 6.051$_{\pm 0.025}$         & 5.681$_{\pm 0.066}$ & G14 \\
\object{AX Cir} & 5.273  & 1.725$_{\pm 0.527}$ & 0.265$_{\pm 0.121}$ & 5.880$_{\pm 0.030}$ & 4.299$_{\pm 0.025}$ & 3.879$_{\pm 0.025}$ & 3.780$_{\pm 0.025}$ & 3.524$_{\pm 0.066}$ & G14 \\
\object{$\delta$ Cep} & 5.366  & 3.364$_{\pm 0.049}$ & 0.075$_{\pm 0.018}$ & 3.950$_{\pm 0.030}$ & 2.683$_{\pm 0.010}$ & 2.396$_{\pm 0.010}$ & 2.294$_{\pm 0.010}$ & 2.104$_{\pm 0.061}$ & B97     \\
\object{CV Mon} & 5.379  & 0.508$_{\pm 0.040}$ & 0.705$_{\pm 0.018}$ & 10.310$_{\pm 0.030}$ & 7.314$_{\pm 0.008}$ & 6.781$_{\pm 0.008}$ & 6.529$_{\pm 0.008}$  & 6.165$_{\pm 0.061}$ & M11 \\
\object{QZ Nor} & 5.401 ${\star}$  & 0.452$_{\pm 0.132}$ & 0.289$_{\pm 0.020}$ & 8.870$_{\pm 0.030}$ & 7.085$_{\pm 0.008}$ & 6.748$_{\pm 0.008}$ & 6.614$_{\pm 0.008}$ & 6.360$_{\pm 0.061}$ & L92      \\
\object{V659 Cen} & 5.622  & 1.355$_{\pm 0.448}$ & 0.151$_{\pm 0.034}$ & 6.620$_{\pm 0.030}$ & 5.177$_{\pm 0.025}$ & 4.907$_{\pm 0.025}$ & 4.651$_{\pm 0.025}$ & 4.583$_{\pm 0.066}$ & G14  \\
\object{CS Vel} & 5.905  & 0.222$_{\pm 0.050}$ & 0.716$_{\pm 0.027}$ & 11.700$_{\pm 0.030}$ & 8.771$_{\pm 0.008}$ & 8.246$_{\pm 0.008}$ & 8.011$_{\pm 0.008}$         & 7.643$_{\pm 0.061}$ & L92  \\
\object{RS Nor} & 6.198  & 0.449$_{\pm 0.043}$ & 0.577$_{\pm 0.036}$ & 10.000$_{\pm 0.030}$ & 7.412$_{\pm 0.010}$ & 6.794$_{\pm 0.010}$ & 6.683$_{\pm 0.010}$ & 6.249$_{\pm 0.061}$ & SPIPS    \\
\object{X Cru} & 6.220  & 0.609$_{\pm 0.061}$ & 0.294$_{\pm 0.019}$ & 8.400$_{\pm 0.030}$ & 6.521$_{\pm 0.025}$ & 6.125$_{\pm 0.025}$ & 5.935$_{\pm 0.025}$ & 5.717$_{\pm 0.066}$  & G14    \\
\object{AW Per} & 6.464  & 1.046$_{\pm 0.349}$ & 0.479$_{\pm 0.016}$ & 7.480$_{\pm 0.030}$ & 5.213$_{\pm 0.008}$ & 4.832$_{\pm 0.008}$ & 4.657$_{\pm 0.008}$ & 4.354$_{\pm 0.061}$ & M11      \\
\object{U Sgr} & 6.745  & 1.461$_{\pm 0.038}$ & 0.408$_{\pm 0.007}$ & 6.690$_{\pm 0.030}$ & 4.506$_{\pm 0.008}$ & 4.100$_{\pm 0.008}$ & 3.912$_{\pm 0.008}$ & 3.637$_{\pm 0.061}$ & M11               \\
\object{ER Car} & 7.720  & 0.889$_{\pm 0.210}$ & 0.111$_{\pm 0.016}$ & 6.820$_{\pm 0.030}$ & 5.310$_{\pm 0.008}$ & 5.034$_{\pm 0.008}$ & 4.896$_{\pm 0.008}$ & 4.698$_{\pm 0.061}$ & G18      \\
\object{SX Vel} & 9.550  & 0.432$_{\pm 0.086}$ & 0.237$_{\pm 0.014}$ & 8.290$_{\pm 0.030}$ & 6.500$_{\pm 0.008}$ & 6.133$_{\pm 0.008}$ & 5.991$_{\pm 0.008}$ & 5.743$_{\pm 0.061}$ & L92     \\
\object{SY Nor} & 12.646  & 0.414$_{\pm 0.053}$ & 0.611$_{\pm 0.059}$ & 9.500$_{\pm 0.030}$ & 6.574$_{\pm 0.008}$ & 6.105$_{\pm 0.008}$ & 5.865$_{\pm 0.008}$ & 5.504$_{\pm 0.061}$ & G18      \\
\object{RS Pup} & 41.443  & 0.503$_{\pm 0.045}$ & 0.451$_{\pm 0.010}$ & 7.010$_{\pm 0.030}$ & 4.365$_{\pm 0.008}$ & 3.828$_{\pm 0.008}$ & 3.619$_{\pm 0.008}$ & 3.276$_{\pm 0.061}$ & L92      \\
\hline
 & \multicolumn{8}{c}{Sample of open cluster Cepheids}  \\
 \hline
\object{CG Cas} & 4.365  & 0.282$_{\pm 0.041}$ & 0.667$_{\pm 0.009}$ & 11.380$_{\pm 0.030}$ & 8.903$_{\pm 0.025}$ & 8.299$_{\pm 0.025}$ & 8.109$_{\pm 0.025}$ &7.775$_{\pm 0.066}$& G14        \\
\object{EV Sct} & 4.398 ${\star}$  & 0.468$_{\pm 0.035}$ & 0.623$_{\pm 0.015}$ & 10.130$_{\pm 0.030}$ & 7.608$_{\pm 0.008}$ & 7.184$_{\pm 0.008}$ & 7.018$_{\pm 0.008}$ &6.658$_{\pm 0.061}$  & L92 \\
\object{CF Cas} & 4.875  & 0.269$_{\pm 0.041}$ & 0.556$_{\pm 0.021}$ & 11.060$_{\pm 0.030}$ & 8.590$_{\pm 0.008}$ & 8.126$_{\pm 0.008}$ & 7.900$_{\pm 0.008}$ &7.608$_{\pm 0.061}$& M11   \\
\object{CV Mon} & 5.379  & 0.523$_{\pm 0.043}$ & 0.705$_{\pm 0.018}$ & 10.310$_{\pm 0.030}$ & 7.314$_{\pm 0.008}$ & 6.781$_{\pm 0.008}$ & 6.529$_{\pm 0.008}$ &6.165$_{\pm 0.061}$ & M11        \\
\object{QZ Nor} & 5.401 ${\star}$  & 0.443$_{\pm 0.036}$ & 0.289$_{\pm 0.020}$ & 8.870$_{\pm 0.030}$ & 7.085$_{\pm 0.008}$ & 6.748$_{\pm 0.008}$ & 6.614$_{\pm 0.008}$ &6.360$_{\pm 0.061}$ & L92        \\
\object{V Cen} & 5.495  & 1.288$_{\pm 0.023}$ & 0.265$_{\pm 0.016}$ & 6.820$_{\pm 0.030}$ & 5.019$_{\pm 0.008}$ & 4.642$_{\pm 0.008}$ & 4.498$_{\pm 0.008}$ &4.249$_{\pm 0.061}$ & L92                \\
\object{CS Vel} & 5.905  & 0.221$_{\pm 0.042}$ & 0.716$_{\pm 0.027}$ & 11.700$_{\pm 0.030}$ & 8.771$_{\pm 0.008}$ & 8.246$_{\pm 0.008}$ & 8.011$_{\pm 0.008}$ &7.643$_{\pm 0.061}$ & L92        \\
\object{V367 Sct} & 6.293  & 0.467$_{\pm 0.041}$ & 1.145$_{\pm 0.043}$ & 11.610$_{\pm 0.030}$ & 7.605$_{\pm 0.008}$ & 6.955$_{\pm 0.008}$ & 6.651$_{\pm 0.008}$ &6.152$_{\pm 0.061}$& L92        \\
\object{U Sgr} & 6.745  & 1.514$_{\pm 0.024}$ & 0.408$_{\pm 0.007}$ & 6.690$_{\pm 0.030}$ & 4.506$_{\pm 0.008}$ & 4.100$_{\pm 0.008}$ & 3.912$_{\pm 0.008}$ &3.636$_{\pm 0.061}$ & M11               \\
\object{RS Ori} & 7.567  & 0.553$_{\pm 0.027}$ & 0.332$_{\pm 0.010}$ & 8.410$_{\pm 0.030}$ & 6.398$_{\pm 0.008}$ & 6.020$_{\pm 0.008}$ & 5.860$_{\pm 0.008}$ &5.589$_{\pm 0.061}$ & M11                \\
\object{DL Cas} & 8.001  & 0.511$_{\pm 0.031}$ & 0.487$_{\pm 0.005}$ & 8.970$_{\pm 0.030}$ & 6.550$_{\pm 0.008}$ & 6.101$_{\pm 0.008}$ & 5.892$_{\pm 0.008}$ &5.593$_{\pm 0.061}$ & M11        \\
\object{S Nor} & 9.754  & 1.025$_{\pm 0.027}$ & 0.182$_{\pm 0.008}$ & 6.420$_{\pm 0.030}$ & 4.674$_{\pm 0.008}$ & 4.288$_{\pm 0.008}$ & 4.149$_{\pm 0.008}$ &3.905$_{\pm 0.061}$ & L92                \\
\object{TW Nor} & 10.786  & 0.383$_{\pm 0.042}$ & 1.190$_{\pm 0.023}$ & 11.670$_{\pm 0.030}$ & 7.442$_{\pm 0.008}$ & 6.712$_{\pm 0.008}$ & 6.375$_{\pm 0.008}$ &5.865$_{\pm 0.061}$ & L92        \\
\object{V340 Nor} & 11.288  & 0.443$_{\pm 0.036}$ & 0.312$_{\pm 0.009}$ & 8.370$_{\pm 0.030}$ & 6.211$_{\pm 0.008}$ & 5.745$_{\pm 0.008}$ & 5.573$_{\pm 0.008}$ &5.284$_{\pm 0.061}$ & L92        \\
\hline
\end{tabular}
\end{center}
\label{table:final_sample}
\tablebib{(G14) \citet{2014AA...566A..37G}; (L92) \citet{Laney:1992fj}; (M11) \citet{2011ApJS..193...12M}; (2MASS) \citet{2006AJ....131.1163S}; (G18) \citet{2018AA...619A...8G}; (W84) \citet{1984ApJS...54..547W}; (B97) \citet{1997PASP..109..645B}; (SPIPS) Light curve fitting with the SPIPS algorithm \citep{2015AA...584A..80M, TrahinPhD}.}
\tablefoot{($\star$) Cepheid pulsating in the first-overtone mode. In this case the period was converted following the approach described in Sect. \ref{app:pulsmodes}. \\
${(*)}$ The parallaxes presented in this table do not include the parallax zero-point offset term. \\
${(**)}$ $W_H$ apparent magnitudes presented in this table do not include the addition of the CRNL term.}
\end{table*}

In order to determine the phase-averaged magnitudes of the CCs in our sample, we searched them in the catalog assembled by \citet{2018AA...619A...8G}. It is a compilation of mean apparent magnitudes in $J$, $H$, $K$, and $V$ bands in different photometric systems, taken from different sources (see Table \ref{table:final_sample}). \citet{Laney:1992fj} provide NIR magnitudes in the SAAO system, to which we set the uncertainties to 0.008 mag following \citet{2018AA...619A...8G}. For homogeneity we converted them into the 2MASS system using the equations from \citet{2007MNRAS.380.1433K}: 
\[
\begin{array}{l l l l l l l l}
J_{\rm 2MASS} ~= - 0.028   ~+ ~J_{\rm SAAO} ~ - ~ 0.047 (J_{\rm SAAO}-K_{\rm SAAO}), \\
H_{\rm 2MASS} = + 0.014 ~+ ~H_{\rm SAAO},             \\
K_{\rm 2MASS} = - 0.015  ~+ ~K_{\rm SAAO} ~ + ~ 0.177 (H_{\rm SAAO}-K_{\rm SAAO}) \\
 \hspace{1.5cm}    - ~ 0.082 (J_{\rm SAAO}-H_{\rm SAAO})^2.
\end{array}
\]
The magnitudes given by \citet{2011ApJS..193...12M} are in the BIRCAM photometric system, we also adopted uncertainties of 0.008 mag, and the magnitudes taken from \citet{1984ApJS...54..547W} and \citet{1997PASP..109..645B} are in the CIT photometric system, with uncertainties of 0.010 mag following \citet{2018AA...619A...8G}. They were all converted into the 2MASS system using the equations from \citet{2011ApJS..193...12M}: 
\[
\begin{array}{l l l l l l l l}
K_{\rm 2MASS} = K_{\rm BIRCAM} + 0.008 ~(J_{\rm BIRCAM}-K_{\rm BIRCAM}) - 0.042,  \\
J_{\rm 2MASS} ~= K_{\rm 2MASS} + 1.052 ~(J_{\rm BIRCAM}-K_{\rm BIRCAM}) - 0.002,   \\
H_{\rm 2MASS} = K_{\rm 2MASS} + 0.993 ~(H_{\rm BIRCAM}-K_{\rm BIRCAM}) + 0.050,   \\
\end{array}
\]
and
\[
\begin{array}{l l l l l l l l}
K_{\rm 2MASS} = K_{\rm CIT} + 0.001~ (J_{\rm CIT}-K_{\rm CIT}) - 0.019,  \\
J_{\rm 2MASS} ~= K_{\rm 2MASS} + 1.068 ~(J_{\rm CIT}-K_{\rm CIT}) - 0.020,   \\
H_{\rm 2MASS} = K_{\rm 2MASS} + 1.000 ~(H_{\rm CIT}-K_{\rm CIT}) + 0.034.   \\
\end{array}
\]
The NIR magnitudes from \citet{2014AA...566A..37G} are derived by template fitting and provided in the 2MASS system. For the remaining stars the mean magnitude is computed as the median of the available data in \citet{1984ApJS...54..547W}, \citet{1992AJ....104.1930S}, and 2MASS \citep{2006AJ....131.1163S}. For RS Nor, the averaged NIR magnitudes were derived by fitting the photometric light curves using the SPIPS algorithm \citep{2015AA...584A..80M}. In the $V$ band, all mean magnitudes are provided in the standard Johnson system and taken from \citet{2015AN....336...70M}. An uncertainty of 0.03 mag on those magnitudes is adopted. 

Based on apparent magnitudes, we built the reddening-free Wesenheit magnitudes $m_H^W$ \citep{1982ApJ...253..575M}, which are a combination of HST-band apparent magnitudes defined by \citet{2018ApJ...855..136R} as
\begin{equation}
m_H^W = F160W - R\,(F555W - F814W)
\label{def_WH}
,\end{equation}
where $R=0.386$ is derived from the \citet{1999PASP..111...63F} formulation with $R_V$ = 3.3. 

Different formulations for the extinction law are available in the literature \citep{1979ARAA..17...73S, 1989ApJ...345..245C}. We adopt the \citet{1999PASP..111...63F} formulation with $R_V = 3.3$, which yields $R_J = 0.86$, $R_H = 0.55$, and $R_K = 0.37$. This   allows a direct comparison of our calibration with that of \citet{2016ApJ...826...56R}, based on HST Fine Guidance Sensor (FGS) and HST Wide Field Camera 3 (WFC3) measurements (see Sect. \ref{subsec:comparisonHST} and \ref{subsec:H0}). 

\citet{2018ApJ...861..126R} provides photometric data in the $F160W$, $F555W,$ and $F814W$ bands for 50 MW Cepheids. Using the stars in common between this sample and the \citet{2018AA...619A...8G} catalog, we derive the set of linear transformations between HST system and ground-based apparent magnitudes, with a scatter of 0.06 mag:
\[
\begin{array}{l l l l l l l l}
F160W = H + 0.25\,(J-H) -0.030,  \\
F555W = V + 0.28\,(J-H) +0.020, \\
F814W = V - 0.47\,(V-H) -0.035.  \\
\end{array}
\]

We note that the transformation from ground-based magnitudes into the HST system requires  accounting for the count-rate non-linearity (CRNL) effect \citep{2018ApJ...861..126R}. This bias affects the infrared detectors on WFC3, and has the consequence of decreasing the magnitude of faint stars like extragalactic CCs, compared to bright stars like Milky Way CCs. This correction is performed by adding 0.026 mag to HST $F160W$ apparent magnitudes \citep{2019wfc..rept....1R}.

We account for the width of the instability strip (IS) by adding in quadrature an additional term in the photometry errors listed in Table \ref{table:final_sample}. In the $V$ band, \citet{2006ApJ...652.1133M} find a dispersion of 0.23 mag; an intrinsic width of 0.22 mag is obtained after subtracting the estimated measurement errors. In the $J$ and $H$ bands, \citet{2017ApJ...842...42M} find a scatter of 0.12 mag, which leaves an intrinsic width of 0.11 mag in NIR bands. In the $K_S$ band, \citet{2004AJ....128.2239P} find a scatter of 0.084 mag based on a Large Magellanic Cloud (LMC) study, which leaves 0.07 mag for the width of the IS after subtracting error measurements. Finally, \citet{2019ApJ...876...85R} find a dispersion of 0.075 mag in the $W_H$ band, yielding an intrinsic width of 0.07 mag for the IS.

In order to compute absolute magnitudes, we need to correct apparent magnitudes from interstellar absorption. We take $E(B-V)$ values from the DDO database \citep{1995IBVS.4148....1F}, which is a compilation of various $E(B-V)$ values from the literature determined in the same system. Following \citet{2018AA...619A...8G}, we apply a multiplicative factor of 0.94 to these reddening values.

        \subsection{Pulsation modes}
        \label{app:pulsmodes}

\begin{table}[]
\caption{Pulsation mode of the Cepheids in our sample.}
\centering
\begin{tabular}{l c c c}
\hline
\hline
Cepheid & GDR2 & Literature & Adopted \\
\hline 
AW Per                  & FU    & FU $^{\rm (a, b)}$            & FU \\
AX Cir                  & FU            & FU $^{\rm (a, b)}$    & FU \\
BP Cir                  & FO    & FU $^{\rm (b)}$, FO $^{\rm (a,c,d,e)}$ & $\star$ \\
CF Cas          & FU    & FU $^{\rm (a, b)}$            & FU \\
CG Cas                  & FU    & FU $^{\rm (a, b)}$            & FU \\
CM Sct                  & FU    & FU $^{\rm (a, b)}$            & FU \\
CS Vel                  & FU    & FU $^{\rm (a, b)}$            & FU \\
CV Mon                  & -             & FU $^{\rm (b)}$                       & FU \\
$\delta$ Cep    & -             & FU $^{\rm (b)}$                       & FU \\
DF Cas                  & FU    & FU $^{\rm (a)}$                       & FU \\
DL Cas          & FU    & FU $^{\rm (a, b)}$            & FU \\
DK Vel          & FO    & FU $^{\rm (b)}$, FO $^{\rm (a,c)}$    & $\star$ \\
ER Car                  & FU    & FU $^{\rm (a, b)}$    & FU \\
EV Sct                  & FO    & FO $^{\rm (a, b)}$            & FO \\
QZ Nor                  & FO    & FU $^{\rm (b)}$, FO $^{\rm (a)}$      & FO \\
RS Nor                  & FU    & FU $^{\rm (a, b)}$            & FU \\
RS Ori          &  FO   & FU $^{\rm (a, b)}$            & FU \\
RS Pup                  & FU            & FU $^{\rm (a, b)}$            & FU \\
S Nor           & FU    & FU $^{\rm (a, b)}$            & FU \\
SX Vel                  & FU    & FU $^{\rm (a, b)}$            & FU \\
SY Nor                  & FU    & FU $^{\rm (a, b)}$            & FU \\
TV CMa          & FU    & FU $^{\rm (a, b)}$    & FU \\
TW Nor          & -             & FU $^{\rm (b)}$                       & FU       \\
U Sgr                   & FU    & FU $^{\rm (a, b)}$            & FU \\
V340 Nor                & -             & FU $^{\rm (a, b)}$                    & FU \\
V350 Sgr        & FO            & FU $^{\rm (a, b)} $   & FU \\
V367 Sct                & -             &  FU $^{\rm (f)}$, FO $^{\rm (b)}$         & FU \\
V532 Cyg        & FO    & FU $^{\rm (b)}$, FO $^{\rm (a)}$              & FO \\
V659 Cen        & FU    & FU $^{\rm (b)}$, FO $^{\rm (a, c)}$   & FU  \\
V950 Sco        & FO    & FU $^{\rm (b)}$, FO $^{\rm (a)}$              & FO \\
V Cen           & FU    & FU $^{\rm (a, b)}$            & FU \\
VW Cru                  & FU    & FU $^{\rm (a, b)}$            & FU \\
X Cru                   & FU    & FU $^{\rm (a, b)}$            & FU \\
\hline
\end{tabular}
\label{table:pulsmodes}
\tablebib{(a)  \citet{2019AA...625A..14R}; 
(b) \citet{2018AJ....156..171L}; 
(c) \citet{2004ASPC..316..209Z}; 
(d) \citet{1992AJ....103.1638E}; 
(e) \citet{2014AstL...40..800U}; 
(f) \citet{2013MNRAS.434.2238A}.}
\tablefoot{FU = fundamental; FO = first overtone; $\star$ = excluded because of uncertain pulsation mode.}
\end{table}

The identification of  first-overtone (FO) Cepheids is essential for the Leavitt law calibration. These stars belong to a parallel sequence on the PL plane and their pulsation period can be converted into a fundamentalized period \citep{1997MNRAS.286L...1F, 2016MNRAS.460.2077K}. We reviewed the different pulsation modes found in the literature for the stars in our sample and followed in particular the pulsation modes provided by the reclassification from \citet{2019AA...625A..14R}. 

The pulsation modes for the Cepheids in our sample are displayed in Table \ref{table:pulsmodes}. The second and third column of this table give the pulsation mode provided by the GRD2 catalog and by the literature, respectively. The last column gives the adopted pulsation mode. 

For \object{BP Cir} and \object{DK Vel}, different pulsation modes were found: they are both classified as FO Cepheids by GDR2 and other studies \citep{2004ASPC..316..209Z, 2019AA...625A..14R}, while they are listed as fundamentals by \citet{2018AJ....156..171L}. The two stars are also consistent with fundamental pulsators in the PL plane. Given the disagreement between the different references about the pulsation mode of \object{BP Cir} and \object{DK Vel}, we decided to exclude them from the sample.

In order to establish accurate PL and PW relations without excluding the first overtones, we converted their observed periods $P_{\rm{FO}}$ into the fundamental mode equivalent period $P_{\rm F}$  using the equation by \citet{2016MNRAS.460.2077K}: 
\begin{align*}
\label{FOeq}
P_{\rm FO} / P_{\rm F} = -0.0239_{\pm 0.0031} \log P_{\rm F} - 0.0404_{\pm 0.0035} \, [\rm Fe/H] \\
+ 0.7187_{\pm 0.0017} .
\end{align*}
Field and cluster Cepheids have  similar distributions in the Galactic plane, so they have similar metallicity distributions and both  can be assumed close to solar \citep{2008A&A...488..731R}. The first overtones of the sample have periods $P_{\rm{FO}}$ comprised between 3 and 4 days. In this range of periods, we can approximate the previous equation by the linear relation:
\begin{equation}
P_{\rm F} = 1.4459 \, P_{\rm FO} - 0.0736
.\end{equation}

The conversion of first overtones into fundamentals is listed in Table \ref{5FOs}. The positions of these Cepheids in the PL plane after the transformation are consistent with the distribution of fundamental pulsators. 

Even though converting first overtones into fundamentals may introduce a small uncertainty on periods, we decided to include them in the sample for the calibration of the Leavitt law. The periods obtained after conversion with the relations from \citet{1997MNRAS.286L...1F} and \citet{2016MNRAS.460.2077K} only differ by 0.006 days. \citet{Gallenne2018} find a difference of less than 1\% between an empirical conversion law and a theoretical one. Including the five first overtones of the sample with their modified periods instead of rejecting them introduces only a very small change on the intercept of the PL relation and improves the precision of the fit. 

\begin{table}[]
\caption{Period conversion of first overtones into fundamental pulsators.}
\centering
\begin{tabular}{l c c c}
\hline
\hline
Cepheid & $P_{\rm{FO}}$  & $P_{\rm F}$ \\
\hline
EV Sct          &   3.091  & 4.396   \\
V532 Cyg        &  3.284  & 4.675  \\
V950 Sco        &  3.380  & 4.814  \\
QZ Nor          &  3.786  & 5.401  \\
\hline
\end{tabular}
\label{5FOs}
\end{table}

\section{Results}
\label{sec:results}

\subsection{Calibration of the Leavitt law}
\label{subsec:Leavitt_law}
        
In this section we combine the 22 Cepheid companions with the 14 open cluster Cepheids. Their parameters are listed in Table \ref{table:final_sample}. We found five Cepheids  present in both samples. For these five stars the companion parallax and the cluster parallax agree within $1\sigma$ except for U Sgr, which is at $1.2\sigma$. In order to avoid any correlation between our two sets of parallaxes,  for these five stars we recomputed the \citet{2018AA...618A..93C} cluster parallaxes as the median of all stars parallaxes after excluding the companion. We found our new cluster parallaxes to differ by $0.5 \, \mu$as at most from the original values, so we adopted these new parallax values and considered the two sources of measurement to be independent and non-correlated. For these five Cepheids, both parallax measurements (cluster and companion) are considered independently in the linear fit. 

In order to calibrate the PL relations and the Period--Wesenheit (PW) relations, we used the approach introduced by \citet{1997MNRAS.286L...1F} and \citet{1999ASPC..167...13A} and we computed the Astrometric Based Luminosity (ABL), defined as\begin{equation}
ABL = 10~ ^{0.2 M_{\lambda}} = \varpi ~10~ ^{0.2 m_{\lambda} - 2} 
,\end{equation}
where $M_{\lambda}$ is the absolute magnitude, $m_{\lambda}$ is the dereddened apparent magnitude, and $\varpi$ is the parallax in milliarcseconds. Calibrating the Leavitt Law following this approach is equivalent to determining the coefficients $a$ and $b$ in the equation
\begin{equation}
 ABL = 10^{\, 0.2 \, [a (\log P - \log P_0) + b]} 
 .\end{equation}

\begin{figure}[]
\centering
\includegraphics[height=8cm]{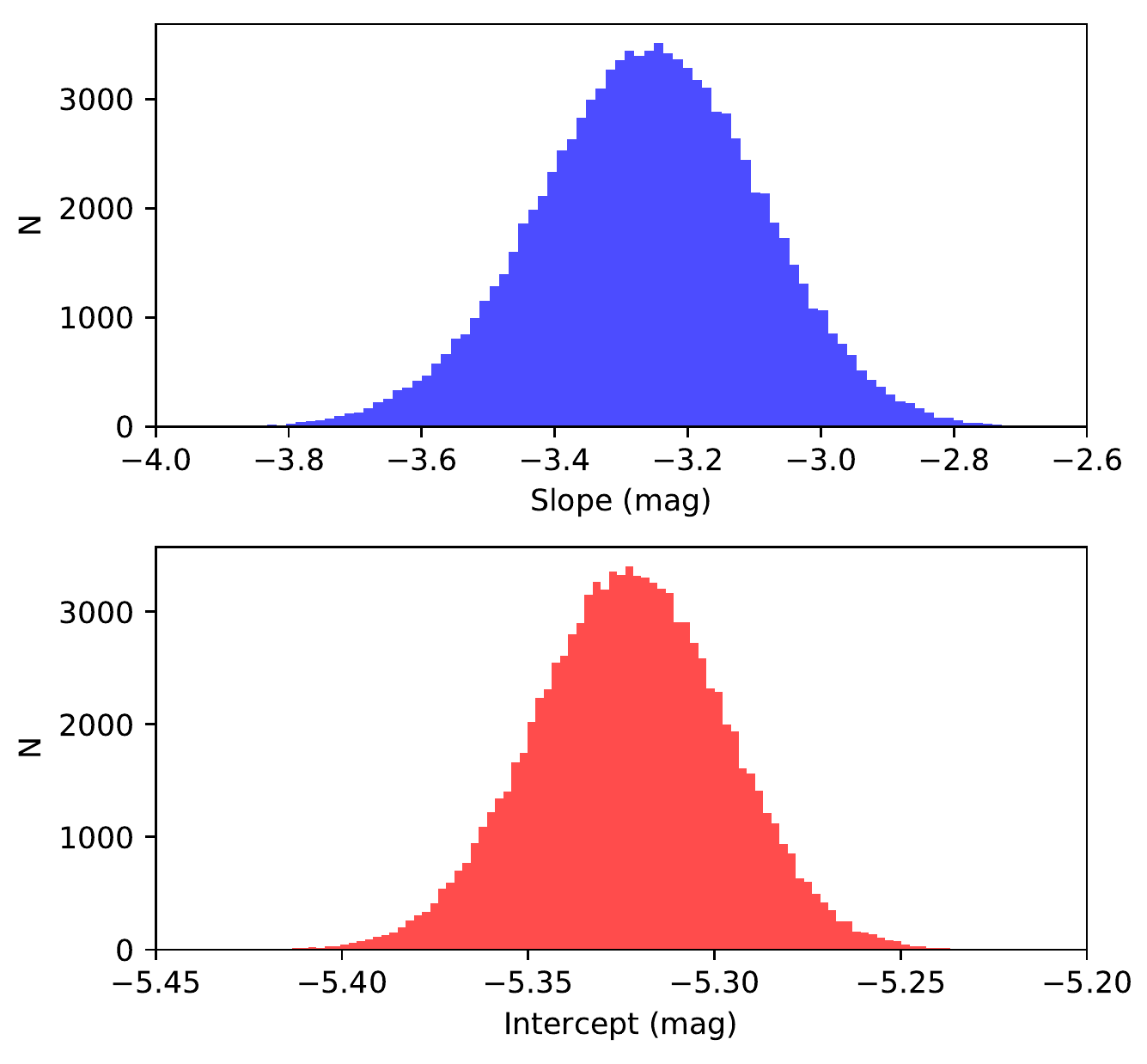}
\caption{Results of the Monte Carlo technique for a PL fit of the form $K_S = a (\log P - \log P_0) + b$ where $\log P_0$ = 0.84. The top and bottom panels respectively show the distribution of the slope $a$ and intercept $b$.}
\label{histogram}
\end{figure}

We performed a weighted fit of the ABL function by using the \texttt{curve\_fit} function from the python Scipy library. The robustness of the fit and of the uncertainties is ensured by a Monte Carlo approach, applied with 100 000 iterations. The distributions of the slope and zero-point of our $K_S$ Leavitt law obtained by this technique are displayed via   histograms in Fig. \ref{histogram}.

\begin{table*}[h]
\caption{Zero-point offset for GDR2 parallaxes found in the literature.}
\centering
\begin{tabular}{l l l c}
\hline
\hline
$\mathrm{ZP}_{\rm GDR2}$  & Reference & Type of sources & Typical $G$ \\
\tiny (mas) & & & \tiny(mag) \\
\hline
$-0.029$                                & \citet{2018AA...616A...2L}            & Quasars                         & 19 \\
$-0.031_{\,\pm 0.011}$  & \citet{Graczyk_2019_02_0}             & Eclipsing binaries                & 9  \\
$-0.0319_{\,\pm 0.0008}$        & \citet{2018AA...616A..17A}            & MW Cepheids             & 8   \\
$-0.035_{\,\pm 0.016}$  & \citet{2018MNRAS.481L.125S}   &  Dwarf stars                  & 9   \\
$-0.041_{\,\pm 0.010}$  & \citet{2019MNRAS.tmp.1036H}   &  Red giants                   & 13   \\
$-0.046_{\,\pm 0.013}$  & \citet{2018ApJ...861..126R}           &  MW Cepheids          & 9  \\
$-0.049_{\,\pm 0.018}$  & \citet{2018AA...619A...8G}            &  MW Cepheids (HST)   & 8  \\
$-0.053_{\,\pm 0.003}$  & \citet{2019ApJ...878..136Z}           & Red giants                    & 13  \\
$-0.054_{\,\pm 0.006}$  & \citet{2019MNRAS.487.3568S}   &  GDR2 RV                      & 12  \\
$-0.057_{\,\pm 0.003}$  & \citet{2018MNRAS.481.1195M}   & RR Lyrae                      & 12 \\
$-0.070_{\,\pm 0.010}$  & \citet{2019AA...625A..14R}            & LMC Cepheids          & 15  \\
$-0.082_{\,\pm 0.033}$  & \citet{2018ApJ...862...61S}           & Eclipsing binaries                & 9  \\
\hline
\end{tabular}
\label{table:all_ZP}
\end{table*}

\begin{table*}[h]
\caption{Coefficients of the PL relation obtained with GDR2 parallaxes of companions and open clusters (left) and with direct parallaxes of Cepheids (right), for different parallax zero-point offsets. The equations are of the form $M = a~ (\log P - 0.84) + b$, and $\rho$ is the correlation between $a$ and $b$.}
\centering
\begin{tabular}{c | c c c c c |  c c c c c | c }
\hline
\hline
Band & $a$ & $b$  & $\rho$ & $\chi^2_r$ & $\sigma$ & $a$ & $b$  & $\rho$ & $\chi^2_r$ & $\sigma$ & ZP \small{(mas)} \\
\hline
                & \multicolumn{5}{c |}{Parallaxes of companions and open clusters} & \multicolumn{5}{c |}{Parallaxes of Cepheids}  \\
\hline
$V$             & -2.486$_{\pm 0.246}$ & -3.782$_{\pm 0.051}$ & 0.15 & 0.27 & 0.18 & -2.093$_{\pm 0.236}$ & -3.888$_{\pm 0.050}$ & 0.09 & 0.58 & 0.23 & -0.031  \\
$J$             & -3.079$_{\pm 0.187}$ & -4.964$_{\pm 0.032}$ & 0.21 & 0.38 & 0.16 & -2.680$_{\pm 0.154}$ & -5.040$_{\pm 0.033}$ & 0.03 & 0.97 & 0.21 & -0.031  \\
$H$             & -3.223$_{\pm 0.185}$ & -5.263$_{\pm 0.032}$ & 0.21 & 0.34 & 0.16 & -2.799$_{\pm 0.151}$ & -5.331$_{\pm 0.033}$ & 0.02 & 0.91 & 0.21 & -0.031  \\
$K_S$   & -3.268$_{\pm 0.165}$ & -5.363$_{\pm 0.026}$ & 0.25 & 0.38 & 0.14 & -2.856$_{\pm 0.127}$ & -5.419$_{\pm 0.028}$ & 0.01 & 1.18 & 0.22 & -0.031  \\
$W_H$   & -3.340$_{\pm 0.180}$ & -5.476$_{\pm 0.030}$ & 0.23 & 0.41 & 0.16 & -2.911$_{\pm 0.141}$ & -5.534$_{\pm 0.031}$ & 0.01 & 0.93 & 0.20 & -0.031   \\
\hline
$V$             & -2.481$_{\pm 0.244}$ & -3.731$_{\pm 0.050}$ & 0.15 & 0.29 & 0.18 & -2.111$_{\pm 0.236}$ & -3.829$_{\pm 0.050}$ & 0.09 & 0.53 & 0.21 & -0.046 \\
$J$             & -3.068$_{\pm 0.184}$ & -4.918$_{\pm 0.032}$ & 0.21 & 0.43 & 0.16 & -2.692$_{\pm 0.153}$ & -4.987$_{\pm 0.033}$ & 0.03 & 0.90 & 0.19 & -0.046  \\
$H$             & -3.215$_{\pm 0.185}$ & -5.217$_{\pm 0.031}$ & 0.22 & 0.37 & 0.16 & -2.811$_{\pm 0.151}$ & -5.278$_{\pm 0.033}$ & 0.03 & 0.82 & 0.19 & -0.046  \\
$K_S$   & -3.257$_{\pm 0.163}$ & -5.323$_{\pm 0.026}$ & 0.25 & 0.44 & 0.14 & -2.865$_{\pm 0.126}$ & -5.370$_{\pm 0.028}$ & 0.01 & 1.09 & 0.19 & -0.046  \\
$W_H$   & -3.332$_{\pm 0.177}$ & -5.432$_{\pm 0.029}$ & 0.23 & 0.47 & 0.17 & -2.923$_{\pm 0.141}$ & -5.483$_{\pm 0.031}$ & 0.01 & 0.85 & 0.18 & -0.046   \\
\hline
$V$             & -2.475$_{\pm 0.243}$ & -3.680$_{\pm 0.050}$ & 0.15 & 0.32 & 0.19 & -2.130$_{\pm 0.235}$ & -3.771$_{\pm 0.049}$ & 0.10 & 0.50 & 0.21 & -0.061 \\
$J$             & -3.060$_{\pm 0.179}$ & -4.874$_{\pm 0.032}$ & 0.20 & 0.52 & 0.17 & -2.703$_{\pm 0.153}$ & -4.936$_{\pm 0.032}$ & 0.03 & 0.86 & 0.18 & -0.061 \\
$H$             & -3.207$_{\pm 0.183}$ & -5.172$_{\pm 0.031}$ & 0.21 & 0.44 & 0.17 & -2.824$_{\pm 0.151}$ & -5.226$_{\pm 0.032}$ & 0.03 & 0.76 & 0.17 & -0.061 \\
$K_S$   & -3.248$_{\pm 0.162}$ & -5.283$_{\pm 0.025}$ & 0.24 & 0.55 & 0.16 & -2.873$_{\pm 0.125}$ & -5.321$_{\pm 0.027}$ & 0.01 & 1.04 & 0.18 & -0.061 \\
$W_H$   & -3.322$_{\pm 0.175}$ & -5.389$_{\pm 0.029}$ & 0.22 & 0.58 & 0.18 & -2.934$_{\pm 0.140}$ & -5.433$_{\pm 0.030}$ & 0.02 & 0.81 & 0.17 & -0.061  \\
\hline
\end{tabular}
\label{table:PL}
\end{table*}

We used the formalism detailed in \citet{Gallenne_2017_11_0}, i.e., we adopted the  linear parameterization
\begin{equation}
M_{\lambda} = b_\lambda + a_\lambda ~(\log P - \log P_0), 
\end{equation}
where $a_\lambda$ and $b_\lambda$ are respectively the slope and the zero-point of the PL relation. This  parameterization removes the correlation between $a_\lambda$ and $b_\lambda$ and minimizes their respective uncertainties.
The optimum value of $\log P_0$ depends on the dataset (see \citet{Gallenne_2017_11_0} for further details)
\begin{equation}
 \log P_0 = \frac{\left<\log P_i/e_i^2\right>}{\left<1/e_i^2\right>} 
 ,\end{equation}
where $\log P_i$ are the periods of the stars, and $e_i$ are the uncertainties on their parallax; $\left<\right>$ denotes the averaging operator. We find our sample centered around $\log P_0 = 0.84$.

GDR2 parallaxes are subject to a zero-point  (ZP) offset, whose value was studied extensively but is still debated. \citet{2018AA...616A...2L} used quasars ($G \sim 19 \, \rm mag$) to derive that Gaia parallaxes are underestimated by 0.029 mas. \citet{2018AA...616A..17A} finds a zero-point of $-0.0319\, \rm mas$ based on Milky Way Cepheids ($G \sim 8 \, \rm mag$), in agreement with the $-0.031\,\rm mas$ estimate by \citet{Graczyk_2019_02_0} from detached eclipsing binaries ($G \sim 9 \, \rm mag$) and surface brightness-color relations. Larger values were also found by \citet{2019AA...625A..14R} and \citet{2018ApJ...862...61S}, who find zero-point offsets of $-0.070\,\rm mas$ and $-0.082\, \rm mas$ respectively. Intermediary values were derived by \citet{2018ApJ...861..126R} and \citet{2018AA...619A...8G}, who estimate $-0.046\,\rm mas$ and $-0.049\,\rm mas$ respectively. The recent determinations of $\mathrm{ZP}_{\rm GDR2}$ are listed in Table \ref{table:all_ZP}. In the following, we adopt ZP$_{\rm GDR2} = -0.046 \, \rm mas$ \citep{2018ApJ...861..126R} from Cepheids, which is close to the median of all values (see Table \ref{table:all_ZP}).

\begin{figure*}[]
\centering
\includegraphics[height=6.1cm]{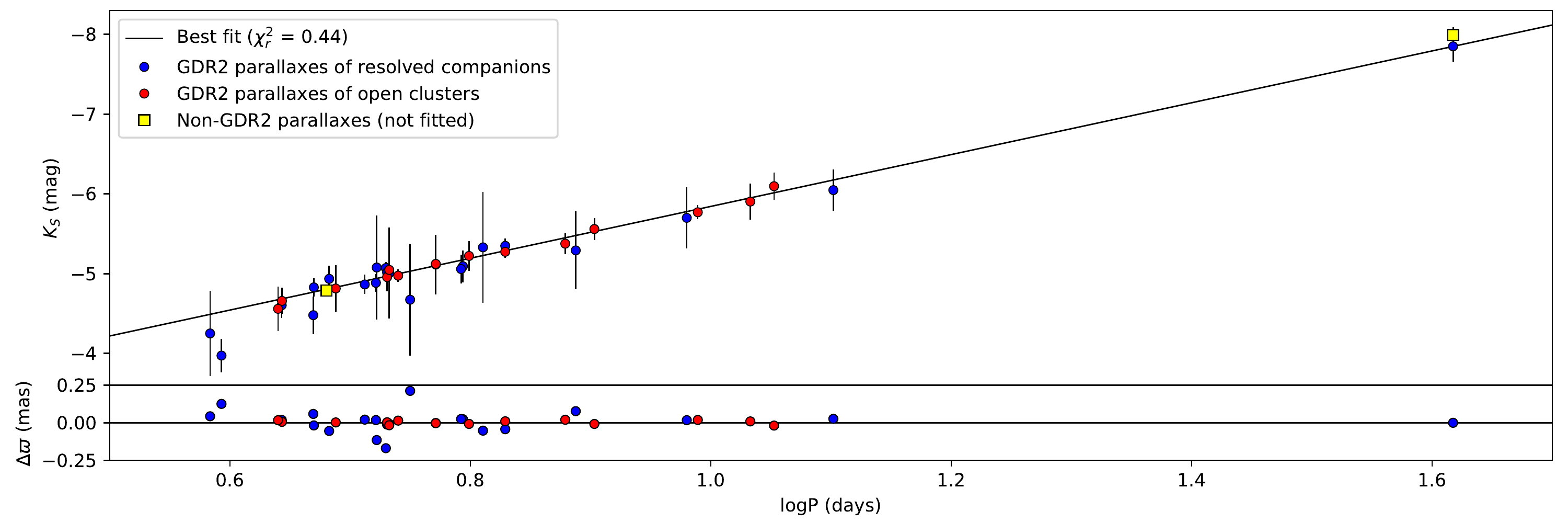}
\caption{Period--luminosity diagram in the $K_S$ band calibrated with GDR2 parallaxes of Cepheids companions (blue) and open clusters (red). The two yellow squares are V1334 Cyg and RS Pup; they are not included in the fit of the PL relation.}
\label{fig:PL_K}
\end{figure*}

The PL coefficients obtained in different bands are listed in Table \ref{table:PL} for different $\mathrm{ZP}_{\rm GDR2}$ values. The Leavitt law calibration in the $K_S$ band is displayed in Fig. \ref{fig:PL_K}. The lower panel shows residuals in terms of parallax, computed as the difference between the input parallax and the parallax given by the best fit. This calibration gives a reduced $\chi^2$ of 0.44 and a dispersion of $\sigma = 0.14$ mag. 

\begin{figure*}[]
\centering
\includegraphics[height=6.1cm]{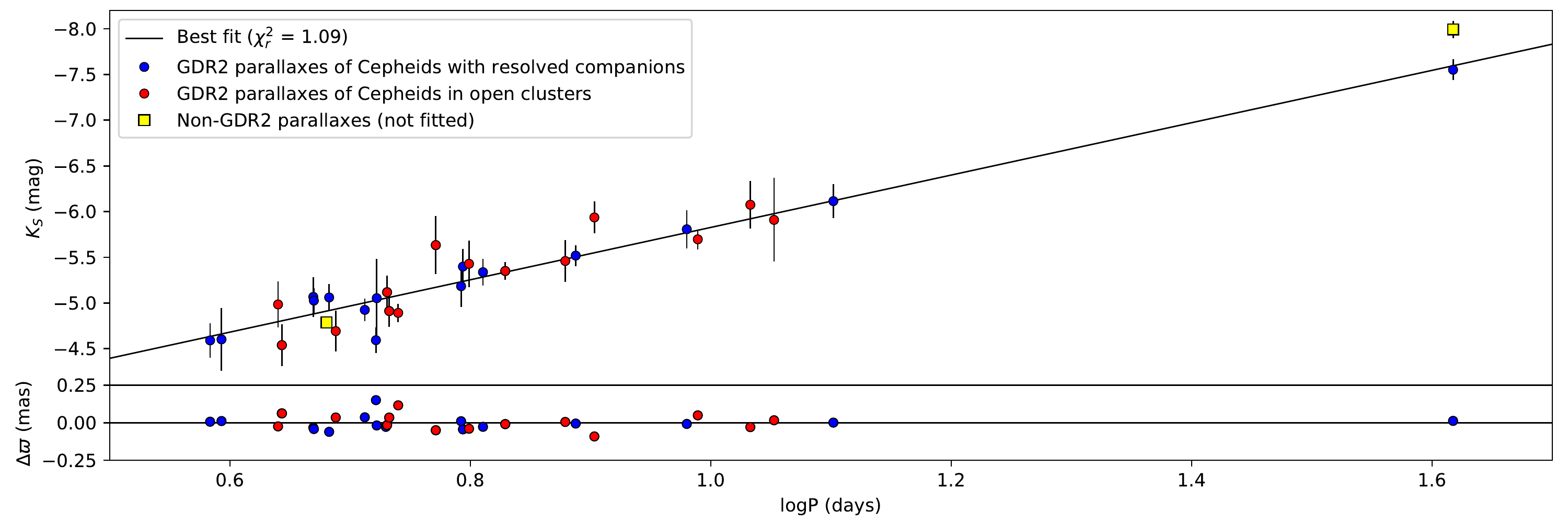}
\caption{Same as Fig. \ref{fig:PL_K}, but using directly GDR2 parallaxes of Cepheids.}
\label{fig:PL_K_cep}
\end{figure*}

An equivalent calibration, based this time on direct Cepheid parallaxes, is presented in Fig. \ref{fig:PL_K_cep}. When the CC parallaxes are adopted, we obtain $\chi^2_r=1.09$ and a dispersion of $\sigma = 0.19$ mag. The dispersion of the PL relation based on Cepheid parallaxes (Fig. \ref{fig:PL_K_cep}) does not appear to be systematic, but rather results in a larger spread not accounted for in the uncertainties. The PL coefficients derived from GDR2 parallaxes of Cepheids are also provided in Table \ref{table:PL}.

We note that very accurate distance measurements are available for a few classical Cepheids, independently of Gaia DR2. They can be used to check the consistency of GDR2 parallaxes. The Cepheid RS Pup has been studied in detail by \citet{Kervella:2014lr} who estimated its parallax to $0.524 \pm 0.022$\,mas using polarimetric HST images of the light echoes propagating in its circumstellar nebula \citep[see also][] {2017A&A...600A.127K}. A second interesting measurement is the distance of the short-period binary Cepheid V1334 Cyg by \citet{Gallenne2018}. It is the most precise parallax determination for a Cepheid, with a value of $1.388 \pm 0.015$ mas. This measurement was obtained by observing the orbit of the system by spectroscopy and optical interferometry. This estimate differs by $3.6 \sigma$ with the GDR2 parallax value ($1.151 \pm 0.066$ mas). These two independent distance measurements are represented by yellow squares on the PL relations in Figs. \ref{fig:PL_K} and \ref{fig:PL_K_cep}, but are not included in the fit since they are not from GDR2. In the case where companion parallaxes and cluster parallaxes are adopted, the two points based on independent measurements agree with the fitted relation at $1 \sigma$. However, in the case of a PL relation based on direct Cepheid parallaxes, both absolute magnitudes derived from the independent points differ by $2.9 \sigma$ from the best fit. The Cepheid RS Pup is particularly interesting since it has a resolved companion listed in our sample (see Table \ref{table:plx_companions}). We note that the RS Pup independent estimate is in very good agreement with the GDR2 parallax of RS Pup companion ($0.503 \pm 0.045$\,mas), but differs by 0.060 mas from the GDR2 parallax of the Cepheid itself (0.584 $\pm$ 0.026 mas).

        \subsection{Comparison with the literature}
        \label{subsec:comparisonHST}
        
In this section, we compare our sample of GDR2 parallaxes with the corresponding parallaxes predicted by a PL calibration based on non-Gaia data. \citet{2016ApJ...826...56R} (hereafter R16) use ten MW Cepheid parallaxes from HST/FGS \citep{2007AJ....133.1810B}, three \textit{Hipparcos} measurements and two Cepheids with parallaxes measured by spatial scanning with the HST/WFC3 \citep{2014ApJ...785..161R, 2016ApJ...825...11C}. These measurements constitute the MW anchor from R16. They combine it with megamasers in $\rm NGC\, 4258$ and eight detached eclipsing binaries in the LMC to derive a final Hubble constant $H_0 = 73.24 \pm 1.74$ \,km\,s$^{-1}$\,Mpc$^{-1}$, associated with the corresponding PL relation in the Wesenheit HST/WFC3 system:
\begin{equation}
M_H^W = -5.93 -3.26\, (\log P -1)
\label{PW_R16}
.\end{equation} 
For the MW anchor only, the $H_0$ value is $76.18 \pm 2.37$ \,km\,s$^{-1}$\,Mpc$^{-1}$. From the ratio of the two $H_0$ values, we offset Eq. \ref{PW_R16} and derive the following PL relation for MW Cepheids only:
\begin{equation}
M_H^W = -5.85 -3.26\, (\log P -1).
\label{PW_R16_MW}
\end{equation} 
We use this Galactic PL calibration based on the Milky Way anchor to compute the predicted parallaxes $\varpi_{R16}$ for each star in our sample:
\begin{equation}
5 \log \varpi_{\rm R16} = M_H^W - m_H^W + 10.
\label{predicted_plx}
\end{equation}
Here $m_H^W$ is the apparent magnitude in the Wesenheit system corrected for the CRNL effect (see Sect. \ref{subsec:photometry}) and $M_H^W$ is derived from the PL relation given by Eq. \ref{PW_R16_MW}. 

\begin{figure}[]
\centering
\includegraphics[height=7.4cm]{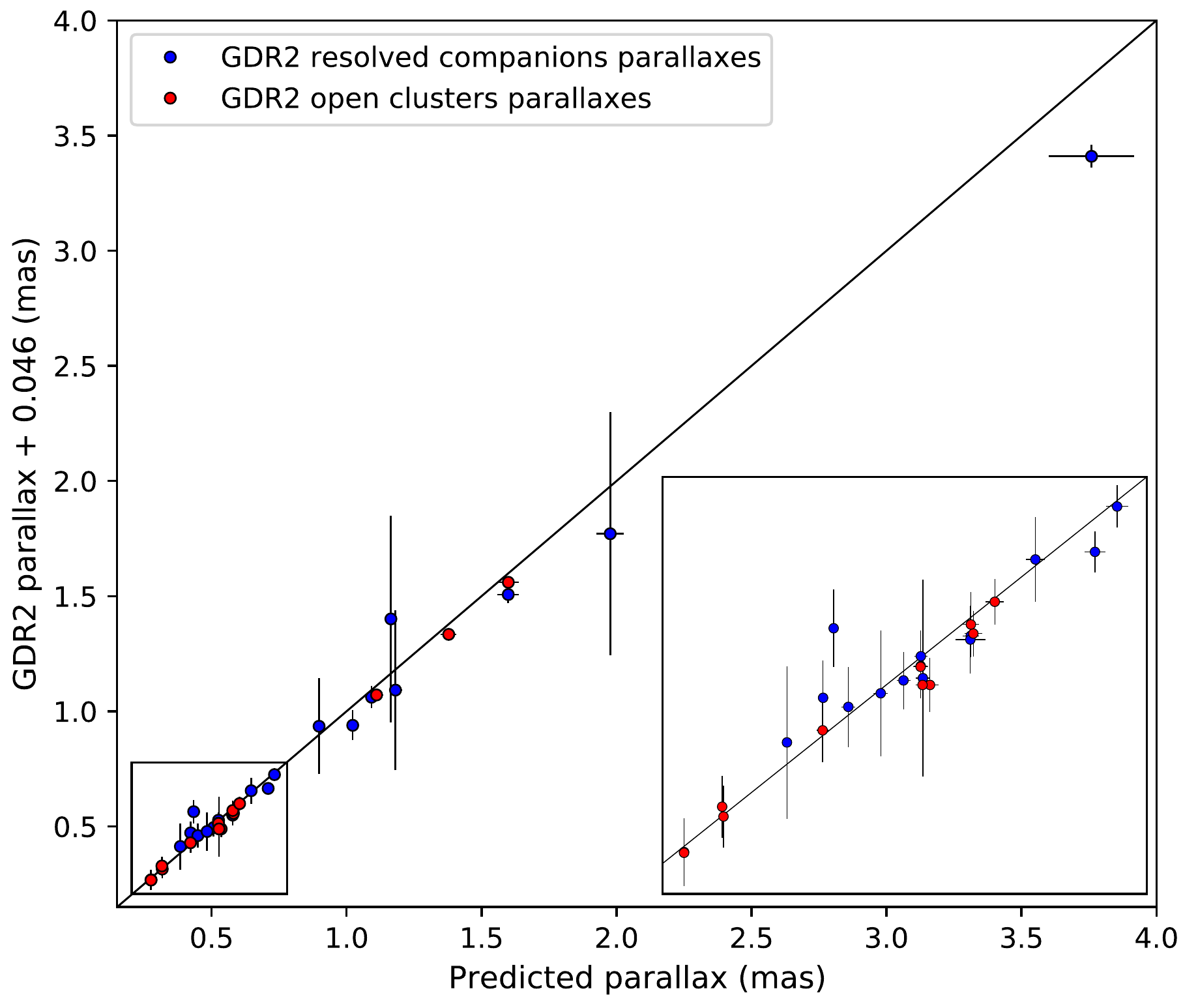}
\caption{Comparison of GDR2 parallaxes of resolved companions and open clusters hosting Cepheids with the predicted parallaxes using the MW PL calibration given in Eq. \ref{PW_R16_MW}. The solid black line corresponds to the identity line.}
\label{fig:predicted_R16}
\end{figure}

The choice of an $R$ value in agreement with \citet{2016ApJ...826...56R} (see Sect. \ref{sec:results}) ensures the consistency of this comparison. To account for the width of the instability strip ($\sigma = 0.07\, \rm mag$ in the $W_H$ band) and for the photometric transformations from ground to HST system ($\sigma = 0.06\, \rm mag$), we set the apparent magnitudes uncertainties to 0.09 mag. Figure \ref{fig:predicted_R16} shows the comparison between the GDR2 parallaxes of our sample of stars corrected by a -0.046 mas offset and the predicted parallaxes from R16. The GDR2 parallaxes appear to be slightly underestimated compared with the predicted values, especially for Cepheids with large parallax values. 

The prototype $\delta$\,Cep is particularly interesting for this study: it hosts a resolved companion with a GDR2 parallax and it is also present in the sample of HST/FGS parallaxes by \citet{2007AJ....133.1810B}. The GDR2 parallax of its companion is $3.393 \pm 0.049$\,mas, while its HST/FGS parallax is $3.66 \pm 0.15$\,mas. These two measurements differ by 1.7$\sigma$ ($7\%$ in relative terms), which agrees with the general trend observed in Fig. \ref{fig:predicted_R16}. We note that $\delta$\,Cep has no valid parallax in GDR2, so its companion parallax is the only possible alternative to HST/FGS measurements.

In Table \ref{comparison_PLs} we present different PL calibrations found in the literature based on various methods and data. \citet{2007AJ....133.1810B} derive a $K$-band PL relation based on HST/FGS parallaxes of seven Galactic Cepheids in the CIT system. We converted this result in the 2MASS system using the relation from \citet{2001AJ....121.2851C}. The investigation by \citet{2007AA...476...73F} provides a PL calibration in the $K_S$ band, based on HST/FGS and Hipparcos parallaxes, as well as infrared surface brightness (IRSB) and interferometric Baade-Wesselink parallaxes. Recently, \citet{2018AA...620A..99G} derived a calibration of the PL relation using a IRSB Baade-Wesselink-type method to determine individual distances to the Cepheids. The result is in the UKIRT system, but the transformation between UKIRT and 2MASS systems given in \citet{2001AJ....121.2851C} shows that this transformation can be neglected. Finally, \citet{2018AA...619A...8G} established a $K_S$-band PL relation based on a large sample of Cepheids parallaxes from GDR2. In Table \ref{comparison_PLs}, we report the coefficients obtained after adopting a GDR2 parallax zero-point of $-0.046\, \rm mas$.

We note that our intercept is very similar to that found by \citet{2018AA...619A...8G}, also based on GDR2 data. However, our calibration shows a significant difference ($\sim$0.1 to 0.2 mag) in intercept with previous calibrations based on HST/FGS data \citep{2007AJ....133.1810B, 2007AA...476...73F}.

\begin{table}[]
\caption{Comparison of our results with other PL relations from the literature. All equations are expressed in the form $K_S = \alpha (\log P -1) + \beta$ in the 2MASS system.}
\centering
\begin{tabular}{l l l }
\hline
\hline
Reference & \hspace{0.6cm} $\alpha$ & \hspace{0.6cm} $\beta$  \\
\hline
\citet{2007AJ....133.1810B}     & $-3.32_{\pm 0.12}$    & $-5.73_{\pm 0.03}$  \\
\citet{2007AA...476...73F}              & $-3.365_{\pm 0.063}$  & $-5.647_{\pm 0.066}$  \\
\citet{2018AA...620A..99G}      & $-3.258_{\pm 0.092}$  & $-5.682_{\pm 0.034}$  \\
\citet{2018AA...619A...8G}              & $-3.028_{\pm 0.067}$  & $-5.867_{\pm 0.087}$  \\
\hline
Present work                            & $-3.257_{\pm 0.163}$  & $-5.844_{\pm 0.037}$  \\
\hline
\end{tabular}
\label{comparison_PLs}
\end{table}

        \subsection{Implications on the distance scale}
        \label{subsec:H0}

The determination of the Hubble constant by \citet{2018arXiv180706209P} exhibit a tension at the $\sim 5\sigma$ level with the latest empirical estimate by \citet{2019ApJ...876...85R} based on LMC Cepheids combined with masers in NGC 4258 and Milky Way parallaxes measured by the HST/FGS, HST/WFC3, and Hipparcos. 

Following the method presented in Section 4 in \citet{2018ApJ...855..136R}, we translate our previous parallax comparison (see Sect. \ref{subsec:comparisonHST}) into a comparison in terms of the Hubble constant. We examine the impact of changing the MW anchor alone on the $H_0$ measurement that depends on three anchors. Therefore, we look at the $H_0$ value from R16 that pertains only to the MW. We use the relation $H_{\rm 0, \,GDR2} = \alpha \, H_{\rm 0, \,R16}$, where $\alpha = \varpi_{\rm GDR2} / \varpi_{\rm R16}$ and $H_{\rm 0, \,R16}$ is the value anchored to Milky Way Cepheids only and is equal to $76.18 \pm 2.37$\,km\,s$^{-1}$\,Mpc$^{-1}$. The expected parallaxes $\varpi_{\rm R16}$ are derived from Eqs. \ref{PW_R16_MW} and \ref{predicted_plx}.

For each star of the sample, we derive the corresponding $\alpha$ value and we adopt a Monte Carlo approach to estimate the final $\alpha$ value averaged over the sample. We performed this calculation on different subsamples and listed the resulting $H_0$ values in Table \ref{table:H0_values}. The uncertainties on $H_0$ include the final error on the R16 estimate excluding the anchors (1.8\%), the error on the estimation of $\alpha$, and finally the uncertainties on the photometric relations to convert ground-based magnitudes into HST magnitudes (1.5\%). Changing the GDR2 parallax offset by 0.015 mas results in a change of 2.6\% in the Hubble constant; therefore, we adopted a confidence interval of 0.015 mas around the -0.046 mas zero-point and added a 2.6\% uncertainty to account for this effect.

\begin{table}[h]
\caption{Hubble constant value derived from the comparison between our GDR2 parallax samples and the predicted parallaxes from R16. The first uncertainties are the statistics combined with the systematics, and the second values account for the effect of the GDR2 parallax zero-point.}
\centering
\begin{tabular}{l c c}
\hline
\hline
                &       $H_0$                                                           & $H_0$                               \\
                &       \tiny (km\,s$^{-1}$\,Mpc$^{-1}$)                                & \tiny (km\,s$^{-1}$\,Mpc$^{-1}$)  \\
\hline
                                &  FU only                                              & FU + FO         \\
\hline
Companions              & $72.83 \pm 2.10 \pm 1.89$                     & $72.49 \pm 2.01 \pm 1.88$  \\
Clusters                        & $73.11 \pm 2.01 \pm 1.90$                     & $73.00 \pm 1.99 \pm 1.90$  \\
\hline
\textbf{All Cepheids}   & \textbf{72.99 $\pm$ 1.89 $\pm$ 1.90}  & \textbf{72.76 $\pm$ 1.86 $\pm$ 1.89} \\
\hline
\end{tabular}
\tablefoot{FU = fundamental mode Cepheids;\\ 
FO = first-overtone mode Cepheids with fundamentalized period.}
\label{table:H0_values}
\end{table}

We obtain a final value of $72.99 \pm 2.68 $ km\,s$^{-1}$\,Mpc$^{-1}$ for fundamental modes only, and  $72.76 \pm 2.65 $ km\,s$^{-1}$\,Mpc$^{-1}$ for all stars included. Both values are very consistent with the LMC and NGC 4258 anchor results derived by \citet{2019ApJ...876...85R}, and also very close to the result by \citet{2019ApJ...886L..27R}. The last value agrees at the 1$\sigma$ level with that of \citet{2020ApJ...891...57F} and at the 2$\sigma$ level with the \citet{2018arXiv180706209P} measurement.

We note that the CCs used to calibrate the PL relation and $H_0$ have lower mean periods than most extragalactic Cepheids found by HST. Though there is no evidence of a break in the PL relation at $\log P=1$ for the Wesenheit magnitude system \citep{1999ApJ...512..711B, 2008ApJ...684..102B, 2016ApJ...826...56R}, it remains important to add longer period Cepheids to the parallax calibration to maintain low systematics.

\section{Conclusions}\label{sec:conclusion}

We presented an original calibration of the Milky Way Leavitt law based on GDR2 parallaxes of resolved Cepheid companions and on GDR2 parallaxes of open clusters hosting Cepheids. Companion and cluster members are not subject to large amplitude photometric and color variability, which reduces the potential for systematic parallax uncertainties. The comparison of our calibration with previous works based on non-Gaia parallaxes indicates a systematic offset between the two measurements. By replacing the trigonometric parallaxes used in R16 by companion and cluster average parallaxes, we render the Milky Way, the LMC, and NGC4258 Leavitt Laws more consistent with one another: we find a MW estimate of $73.0 \pm 2.7 $ km\,s$^{-1}$\,Mpc$^{-1}$ for fundamental modes only and of $H_0 = 72.8 \pm 2.7 $ km\,s$^{-1}$\,Mpc$^{-1}$ for all stars included.

The inclusion of the variability of CCs is not expected in the astrometric processing of the third Gaia data release. However, the effects of the systematics due to the absence of chromaticity correction on Cepheids parallaxes should be reduced in the next releases thanks to the larger number of measurements. The future developments will help to pursue the community goal to measure $H_0$ with utmost precision and accuracy.

\begin{acknowledgements}
We gratefully acknowledge D. Graczyk, S. Borgniet and L. Inno for their comments and corrections. We thank T. J. Calderwood from AAVSO for the photometry of Polaris. The research leading to these results has received funding from the European Research Council (ERC) under the European Union's Horizon 2020 research and innovation programme under grant agreement No 695099 (project CepBin). This work has made use of data from the European Space Agency (ESA) mission Gaia (http://www.cosmos.esa.int/gaia), processed by the Gaia Data Processing and Analysis Consortium (DPAC, http://www.cosmos.esa.int/web/gaia/dpac/consortium). Funding for the DPAC has been provided by national institutions, in particular the institutions participating in the Gaia Multilateral Agreement. The authors acknowledge the support of the French Agence Nationale de la Recherche (ANR), under grant ANR-15-CE31-0012-01 (project UnlockCepheids). W.G. and G.P. gratefully acknowledge financial support for this work from the BASAL Centro de Astrofisica y Tecnologias Afines (CATA) AFB-170002. W.G. acknowledges financial support from the Millenium Institute of Astrophysics (MAS) of the Iniciativa Cientifica Milenio del Ministerio de Economia, Fomento y Turismo de Chile, project IC120009. We acknowledge support from the IdP II 2015 0002 64 and DIR/WK/2018/09 grants of the Polish Ministry of Science and Higher Education and Polish National Science Centre grants MAESTRO UMO-2017/26/A/ST9/00446. This research made use of Astropy7, a community-developed core Python package for Astronomy \citep{2018AJ....156..123A}. We used the SIMBAD and VIZIER databases and catalog access tool at the CDS, Strasbourg (France), and NASA’s Astrophysics Data System Bibliographic Services.
\end{acknowledgements}

\bibliographystyle{aa}
\bibliography{Breuval_AA_2020.bib}

\begin{appendix}

\section{Field charts of open clusters}

\begin{figure*}[]
\centering
\includegraphics[height=8cm]{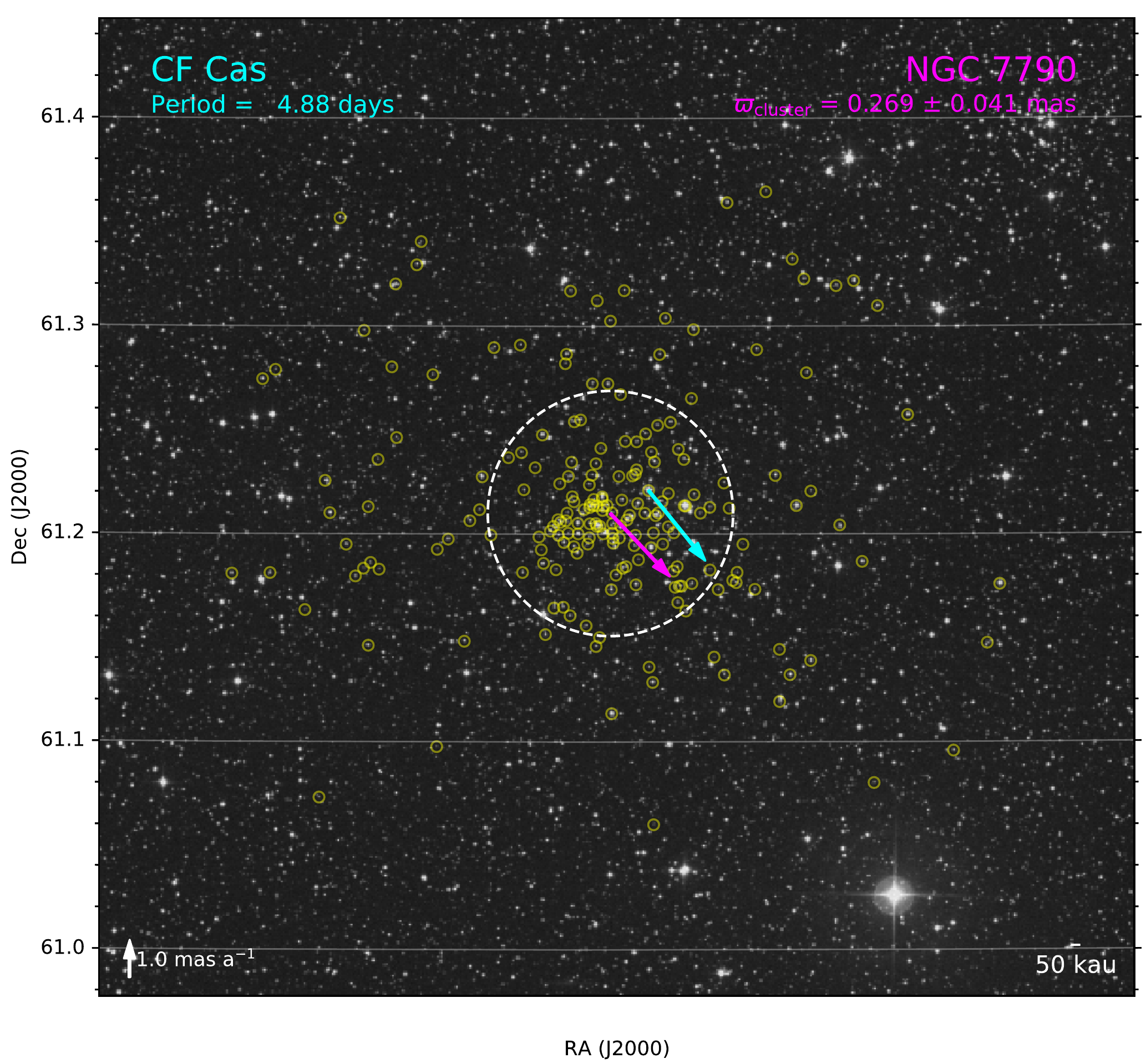}
\includegraphics[height=8cm]{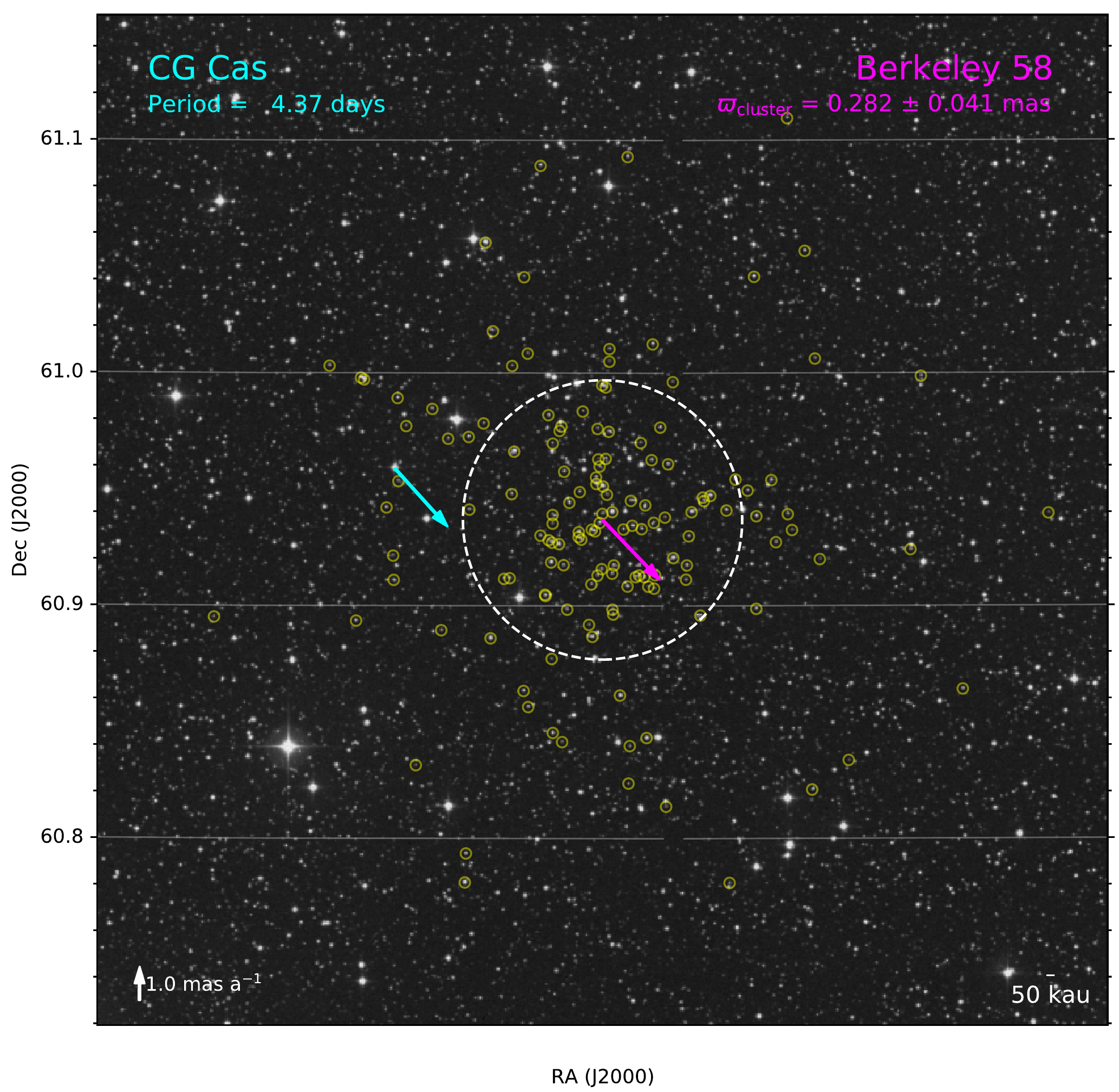}
\includegraphics[height=8cm]{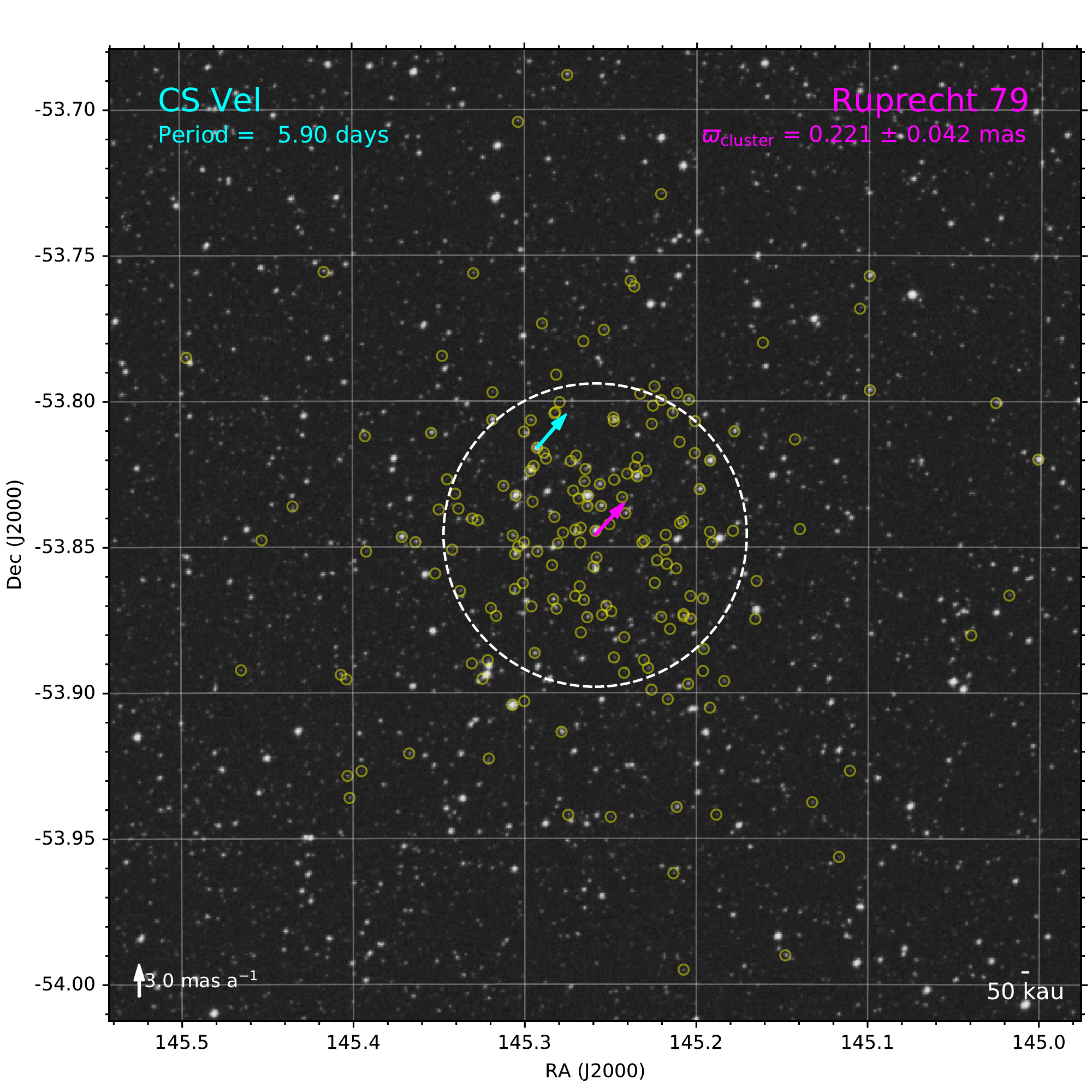}
\includegraphics[height=8cm]{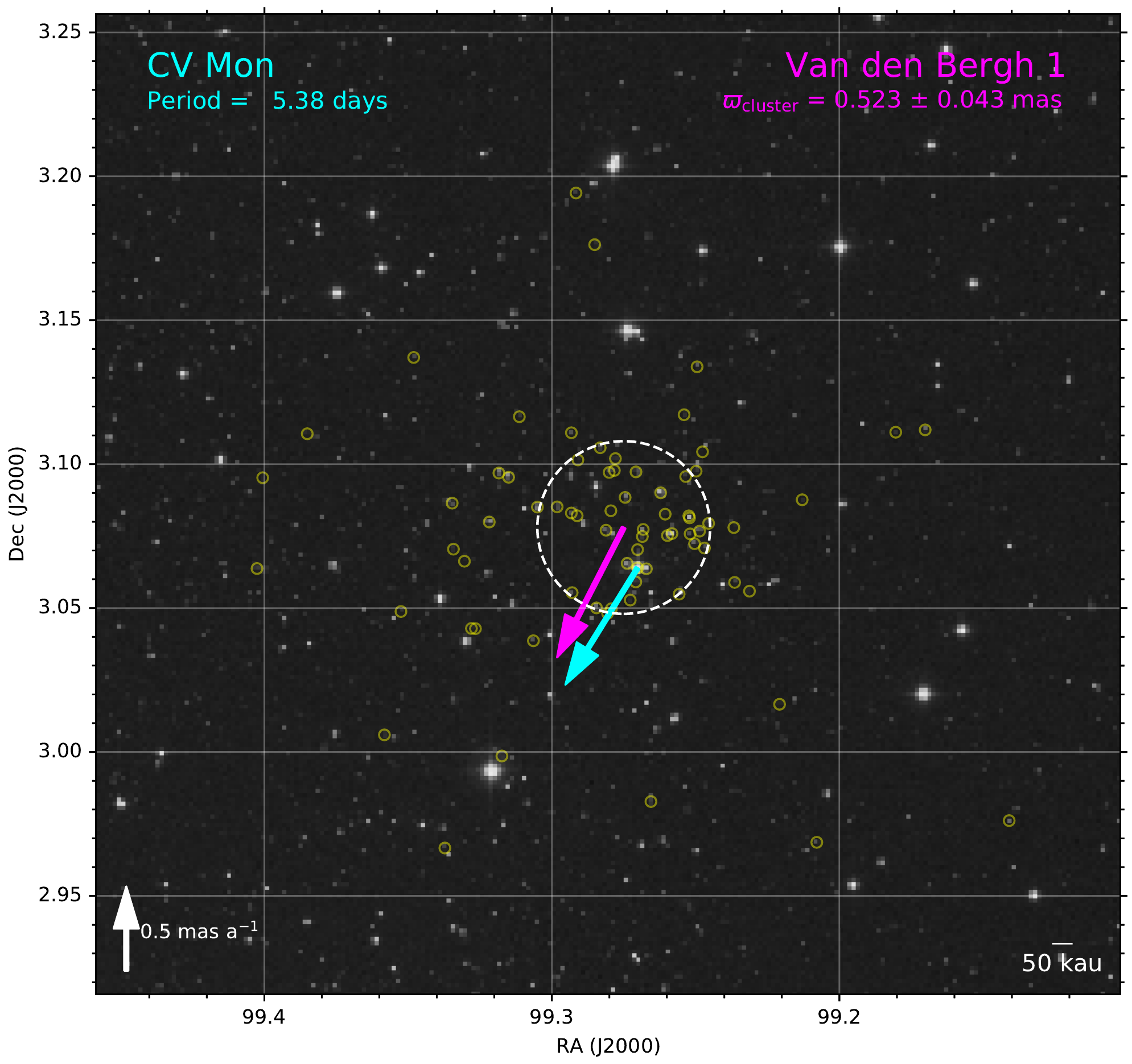}
\includegraphics[height=8cm]{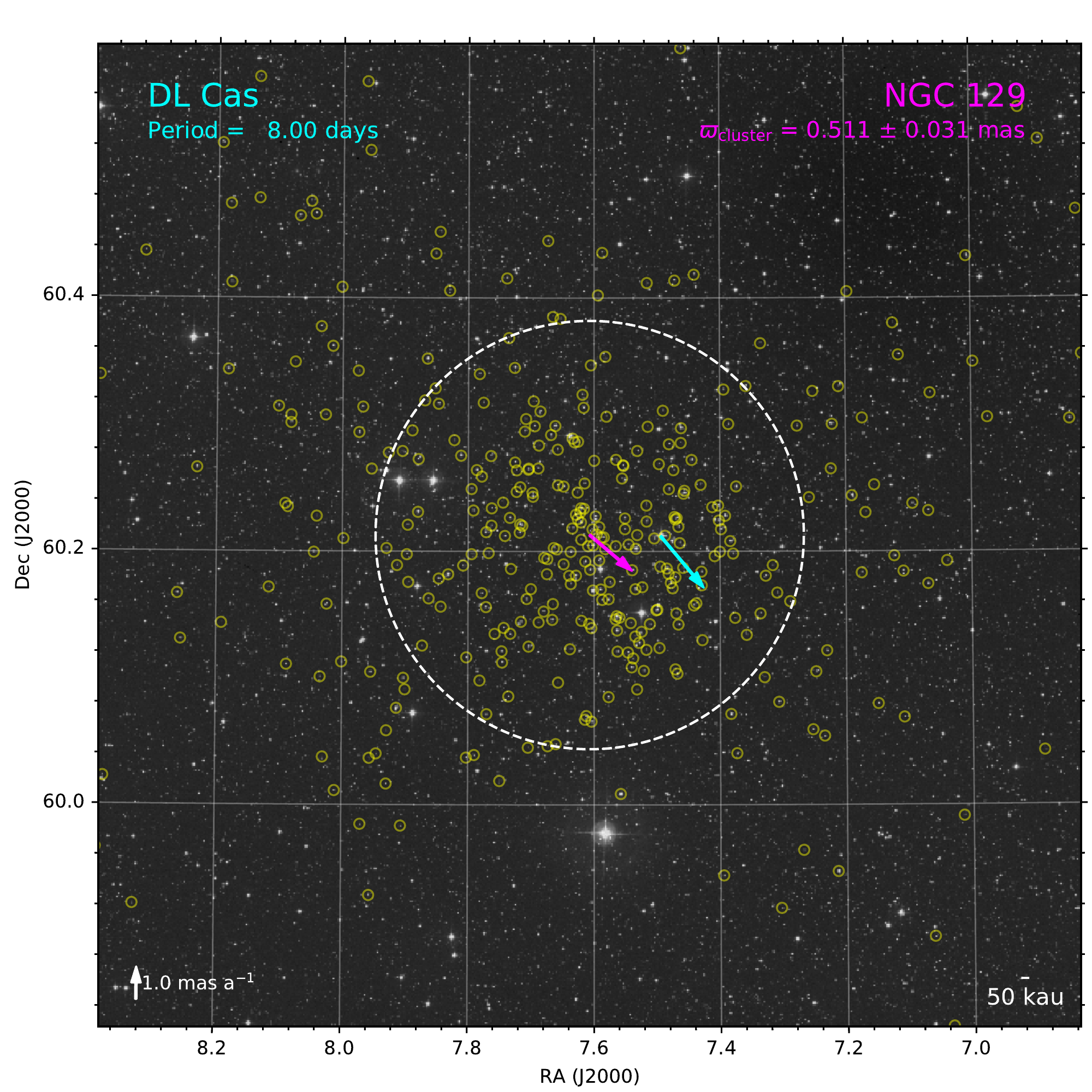}
\includegraphics[height=8cm]{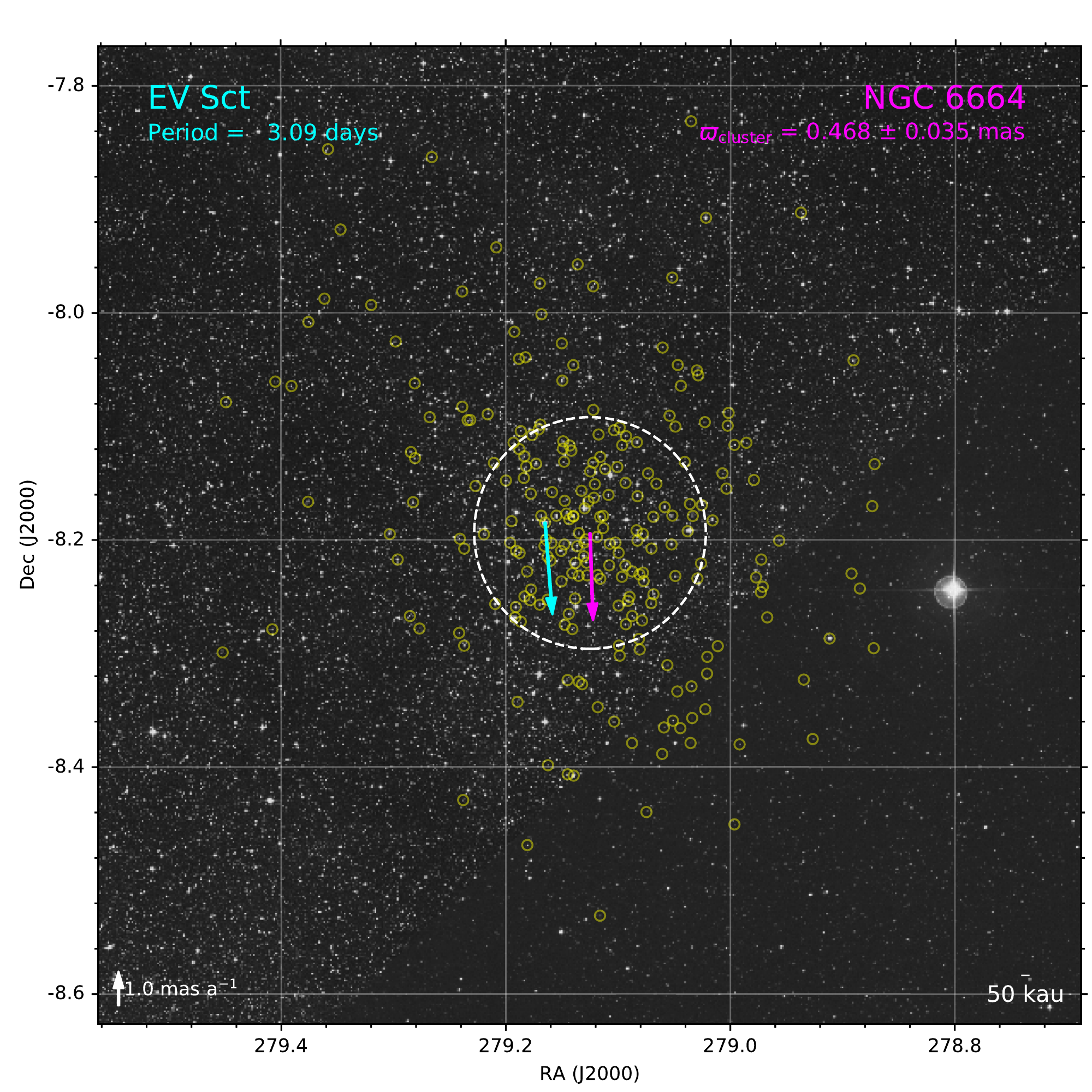}
\caption{Field charts of our candidate Cepheids in their host clusters. The white dashed circles show the radius $r_{50}$ containing half of the cluster stars, and each yellow circle shows a cluster member. The blue and pink arrows show the Cepheid and cluster proper motion, respectively.}
\label{fields_1_of_3}
\end{figure*}

\begin{figure*}[]
\centering
\includegraphics[height=8cm]{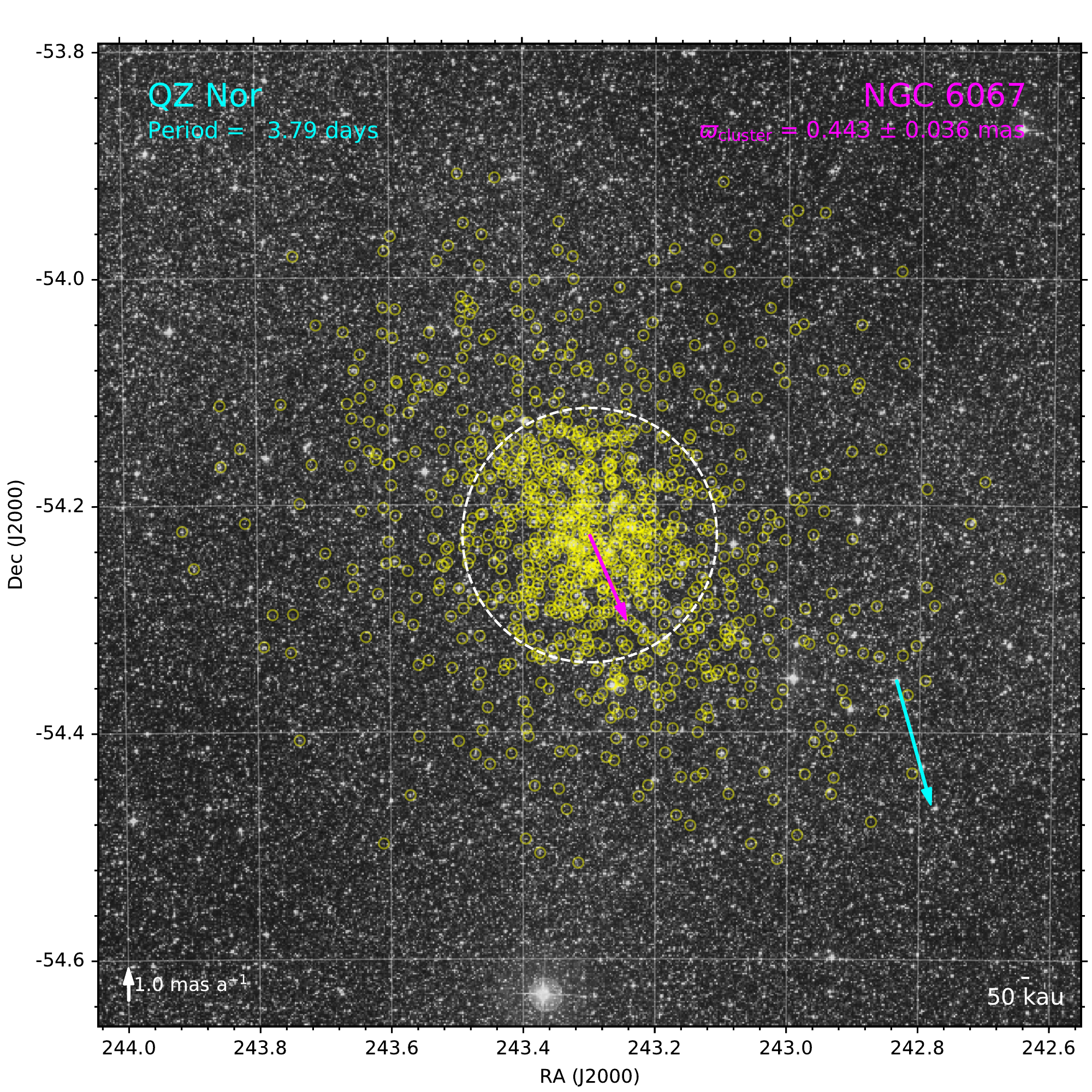}
\includegraphics[height=8cm]{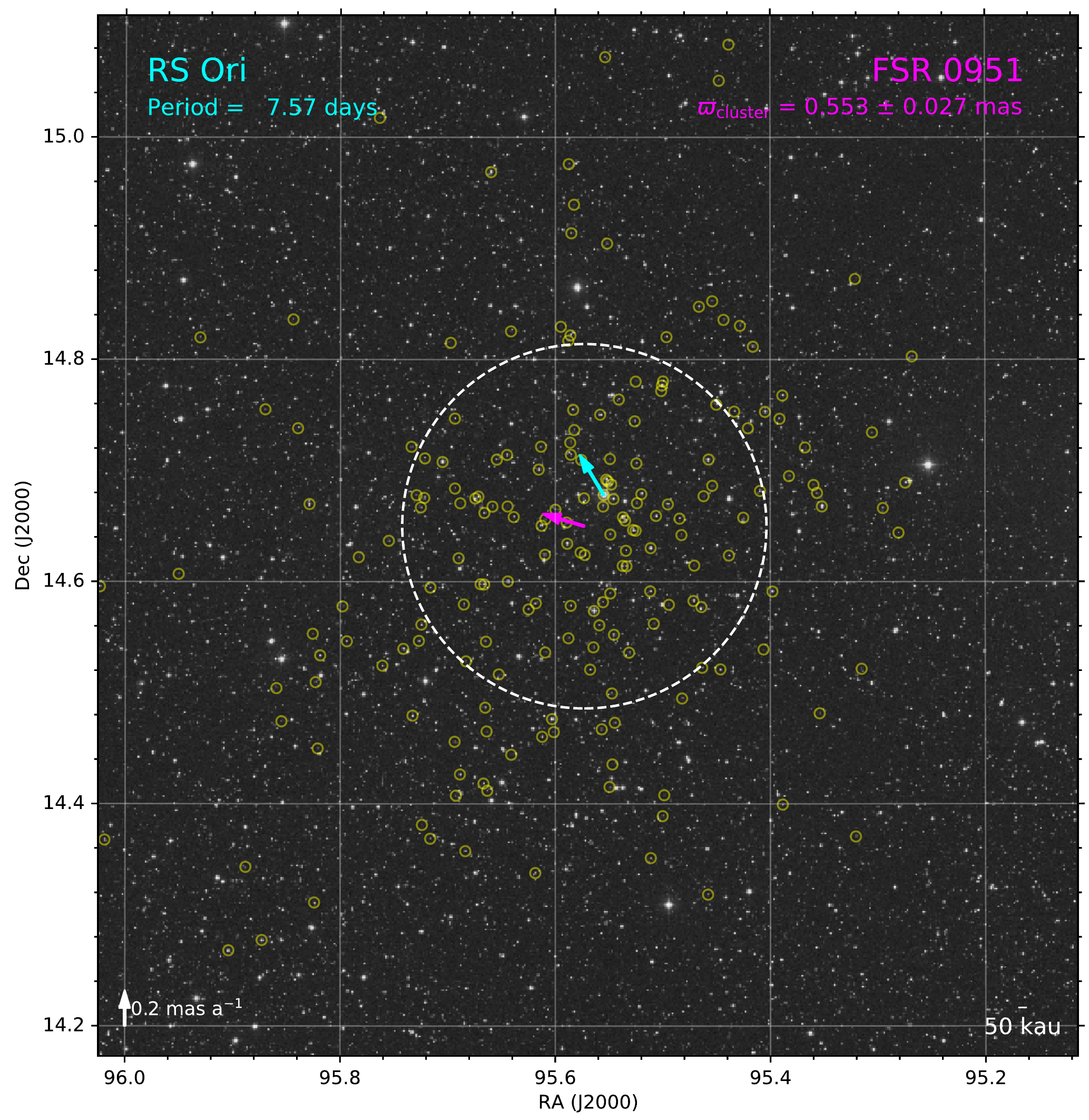}
\includegraphics[width=8.1cm]{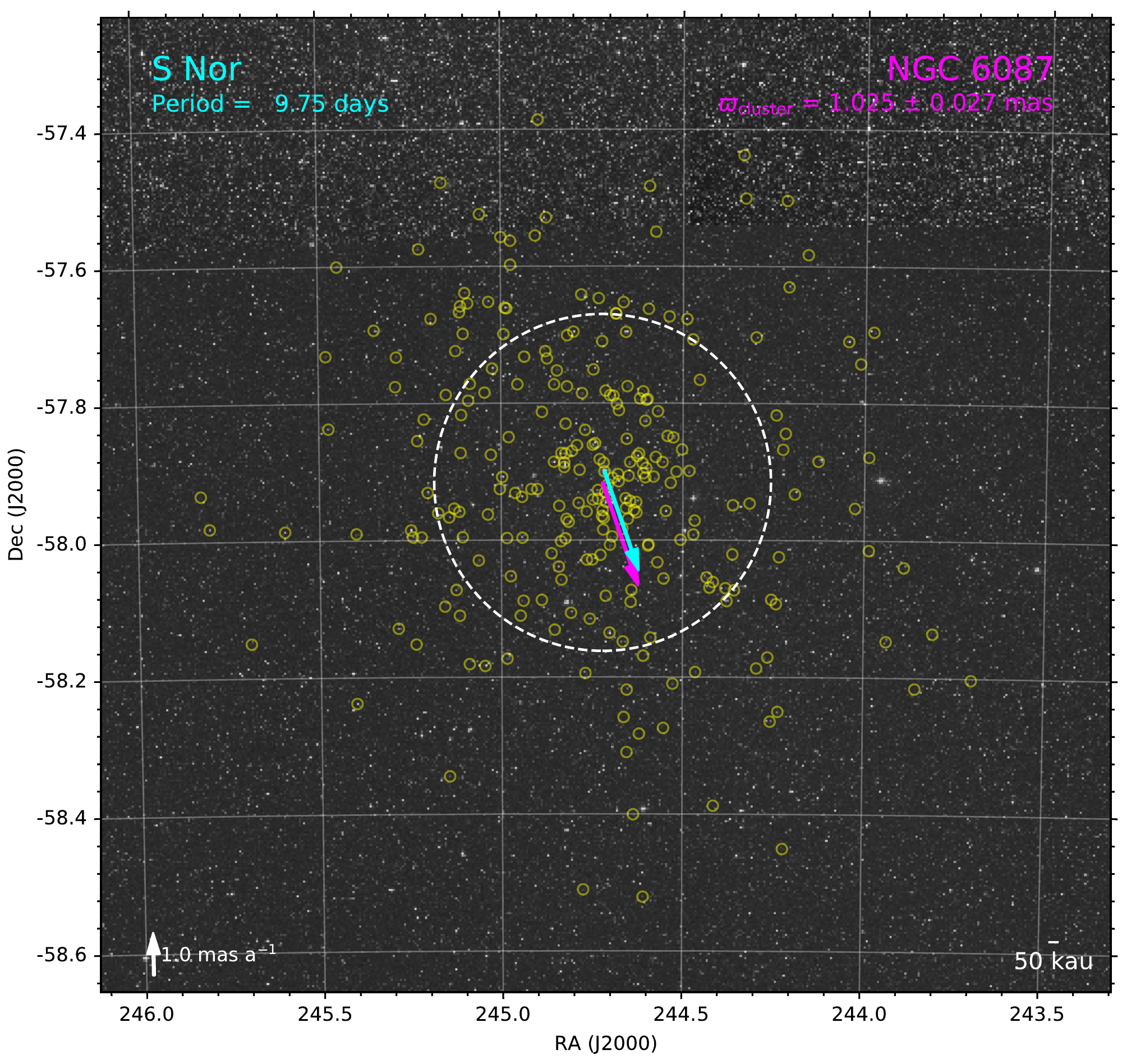}
\includegraphics[width=8.1cm]{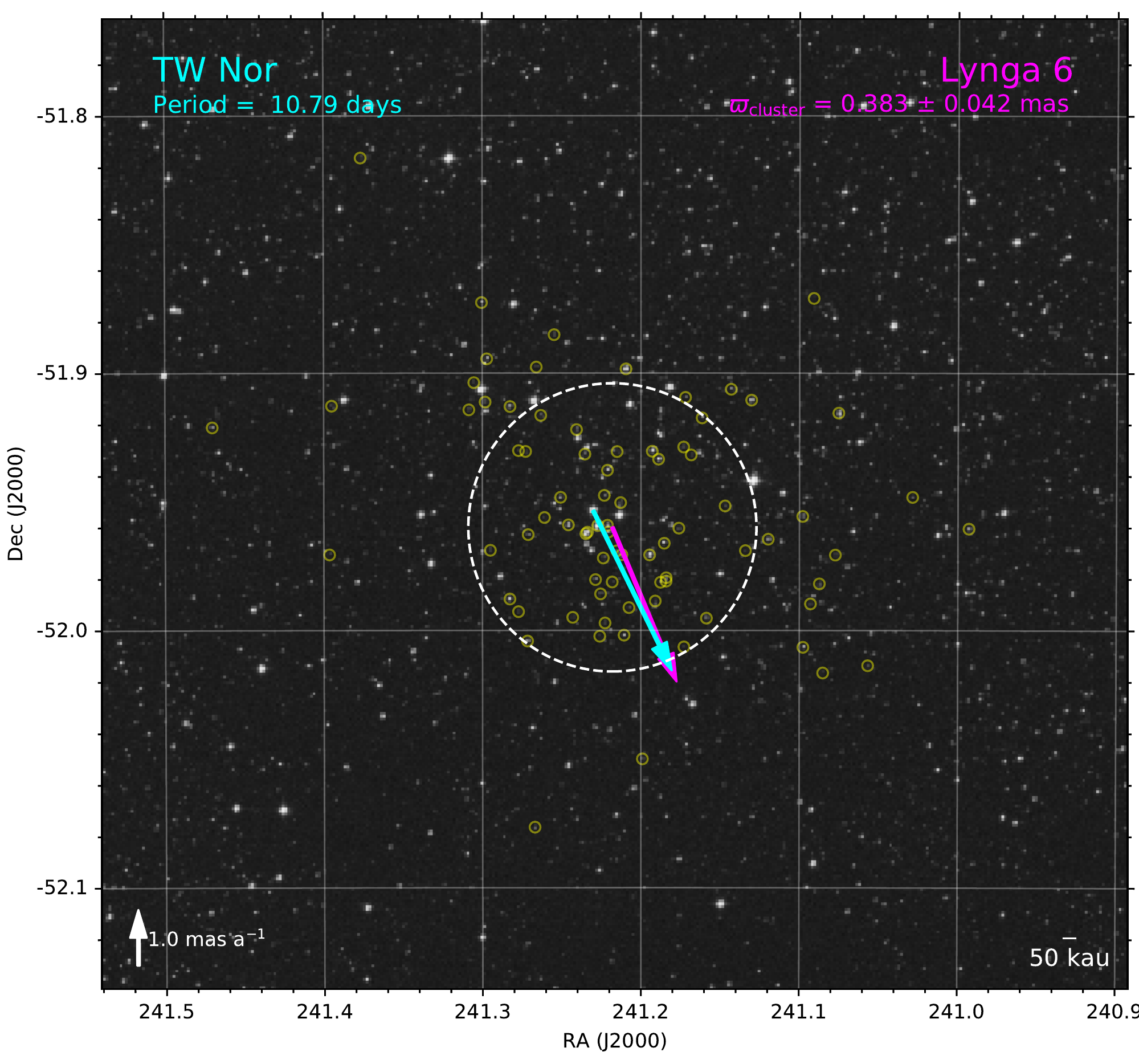}
\includegraphics[height=8cm]{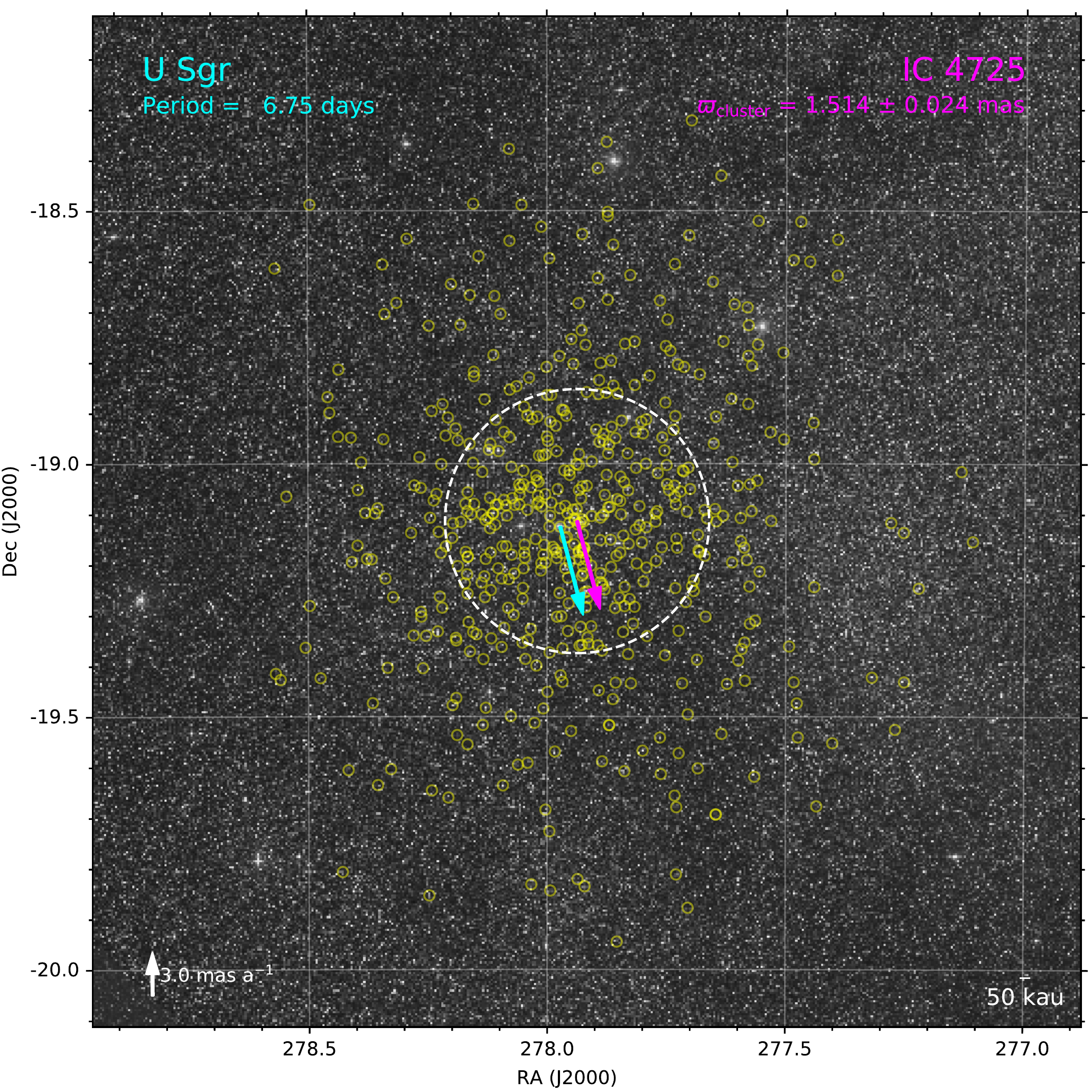}
\includegraphics[height=8cm]{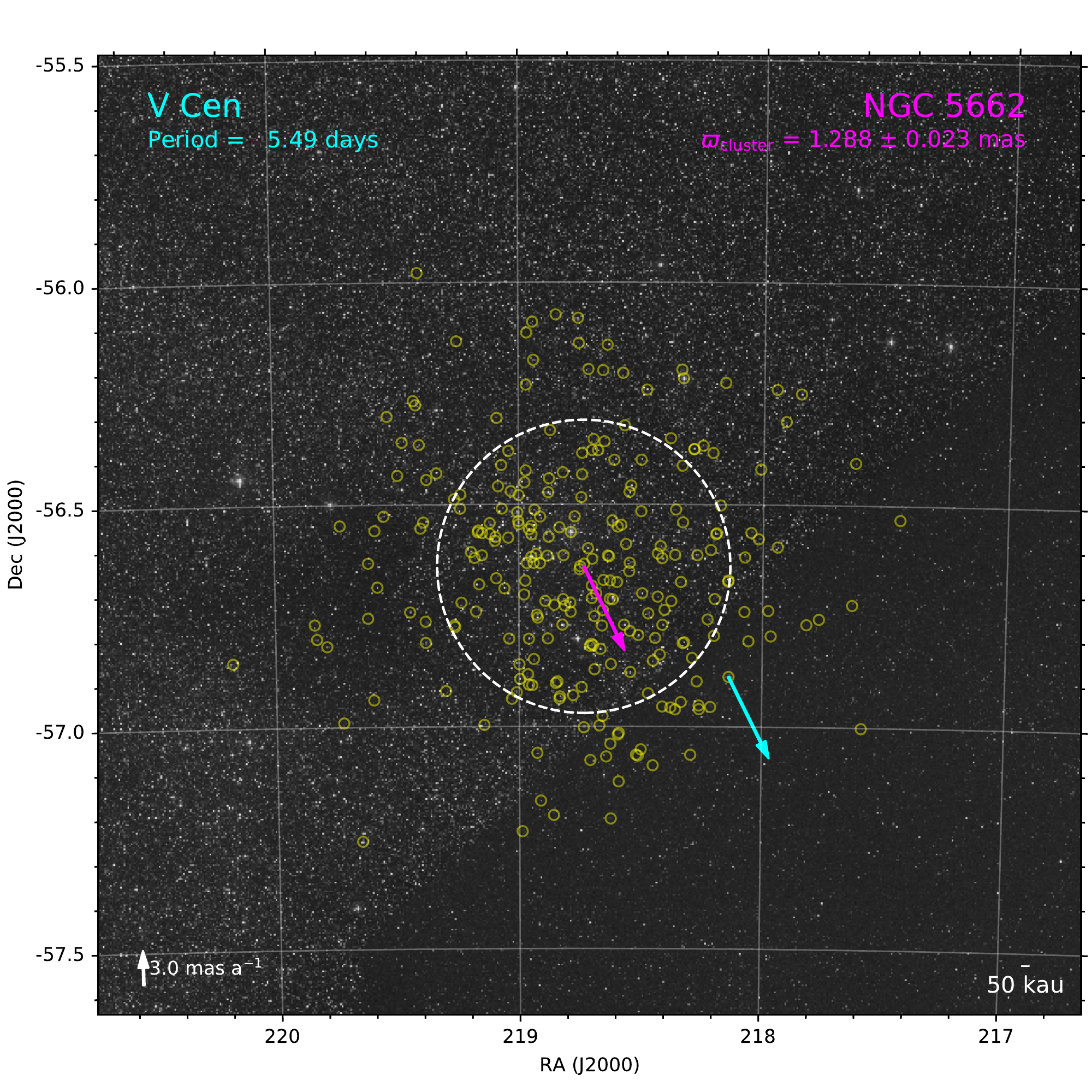}
\caption{Continuation of Fig. \ref{fields_1_of_3}}
\label{fields_2_of_3}
\end{figure*}

\begin{figure*}[]
\centering
\includegraphics[height=8cm]{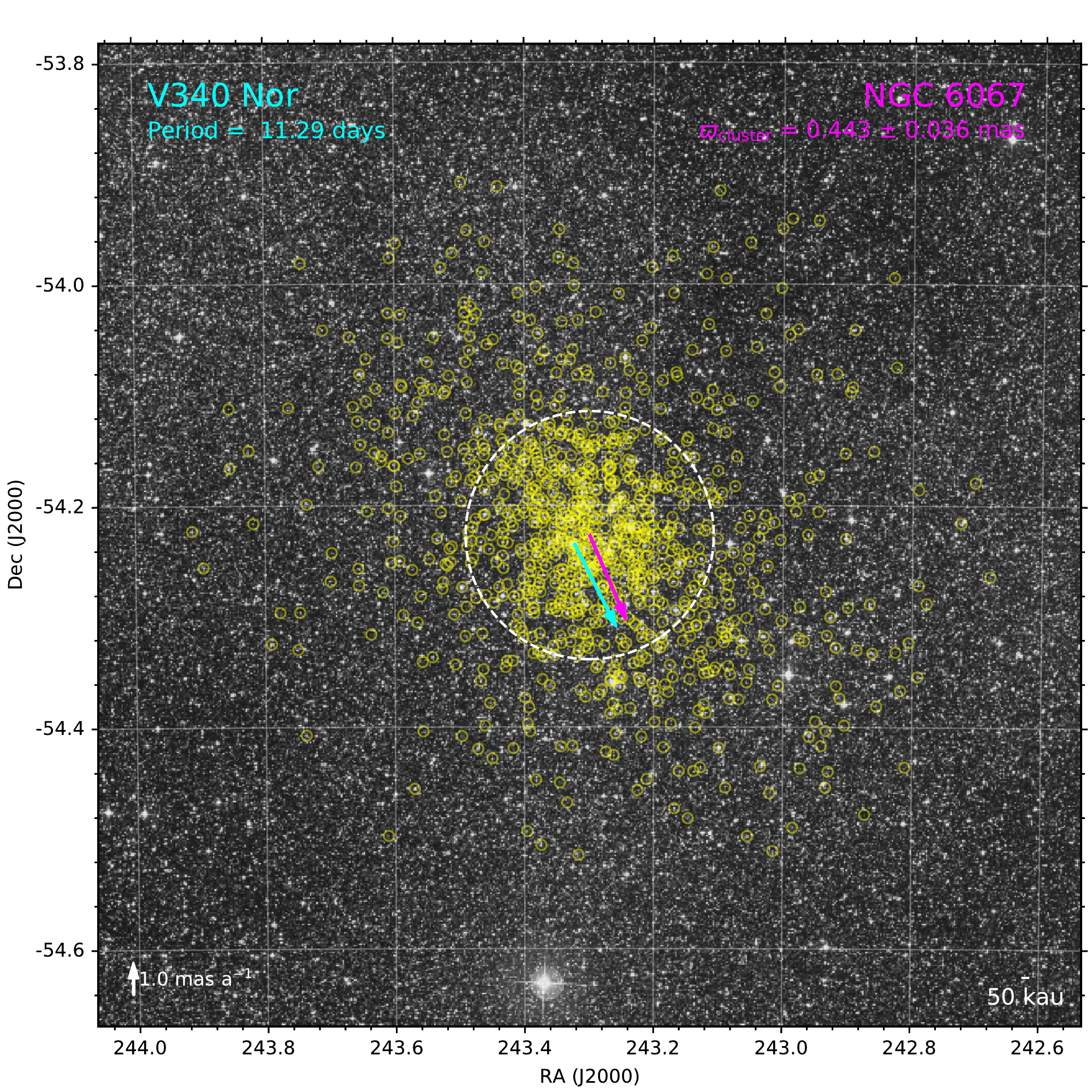}
\includegraphics[height=8cm]{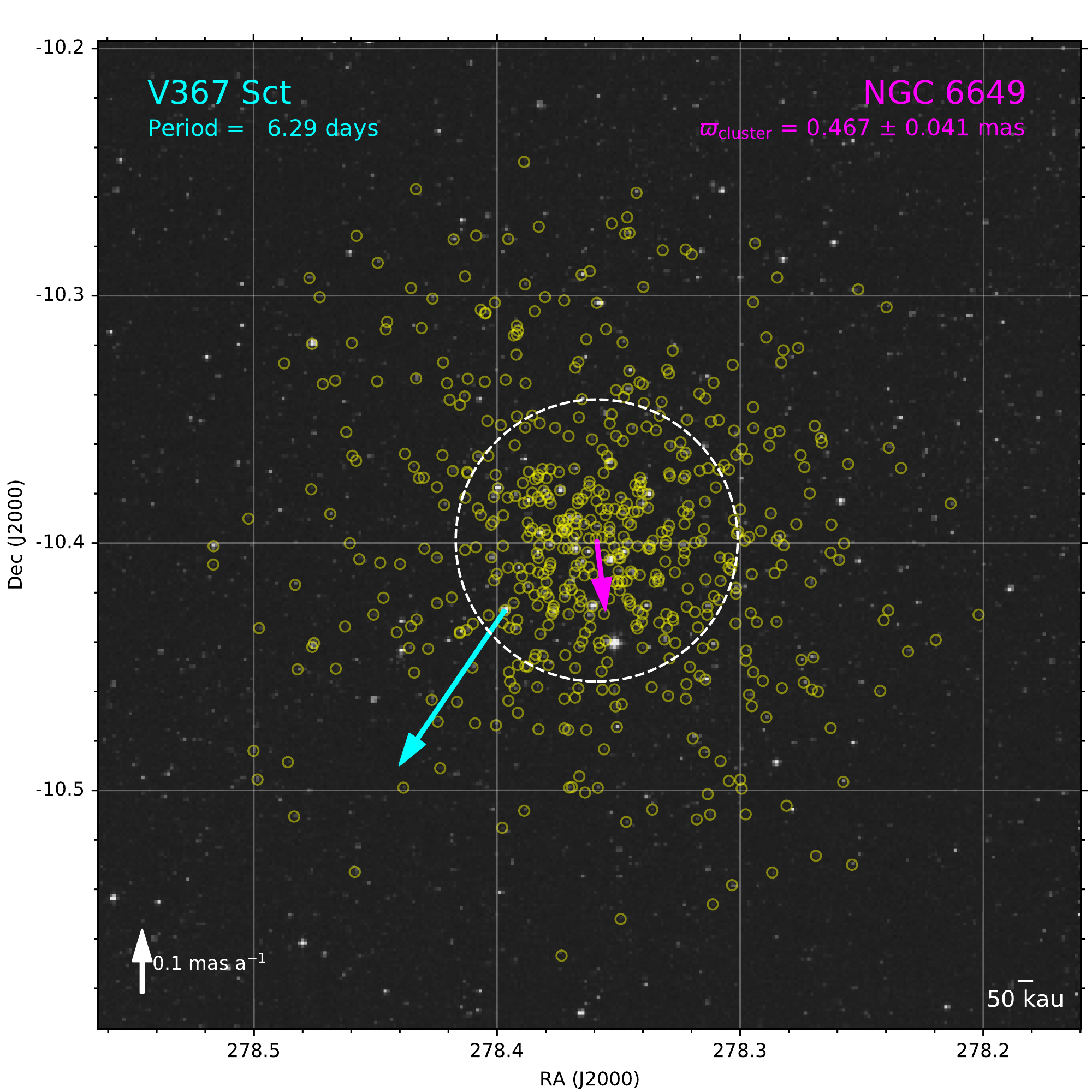}
\caption{Continuation of Fig. \ref{fields_2_of_3}}
\label{fields_3_of_3}
\end{figure*}

\end{appendix}

\end{document}